\documentclass[aip,jap,showpacs,reprint,floatfix,superscriptaddress]{revtex4-1}

\usepackage[utf8]{inputenc}
\usepackage{graphicx}
\usepackage{dcolumn} 
\usepackage{amsmath,amssymb,bm}
\usepackage{esint}
\usepackage[caption=false]{subfig}
\usepackage{color}

\begin{document}
\title{Thermal conductivity of III-V semiconductor superlattices}

\author{S. Mei}
\email{song.mei@wisc.edu}
\author{I. Knezevic}
\email{irena.knezevic@wisc.edu}
\affiliation{Department of Electrical and Computer Engineering, University of Wisconsin-Madison, Madison, WI 53706, USA}
\date{\today}

\begin{abstract}
This paper presents a semiclassical model for the anisotropic thermal transport in III-V semiconductor superlattices (SLs). An effective interface rms roughness is the only adjustable parameter. Thermal transport inside a layer is described by the Boltzmann transport equation in the relaxation time approximation and is affected by the relevant scattering mechanisms (three-phonon, mass-difference, and dopant and electron scattering of phonons), as well as by diffuse scattering from the interfaces captured via an effective interface scattering rate. The in-plane thermal conductivity is obtained from the layer conductivities connected in parallel. The cross-plane thermal conductivity is calculated from the layer thermal conductivities in series with one another and with thermal boundary resistances (TBRs) associated with each interface; the TBRs dominate cross-plane transport. The TBR of each interface is calculated from the transmission coefficient obtained by interpolating between the acoustic mismatch model (AMM) and the diffuse mismatch model (DMM), where the weight of the AMM transmission coefficient is the same wavelength-dependent specularity parameter related to the effective interface rms roughness that is commonly used to describe diffuse interface scattering. The model is applied to multiple III-arsenide superlattices, and the results are in very good agreement with experimental findings. The method is both simple and accurate, easy to implement, and applicable to complicated SL systems, such as the active regions of quantum cascade lasers. It is also valid for other SL material systems with high-quality interfaces and predominantly incoherent phonon transport.
\end{abstract}


\maketitle 
\section{Introduction} \label{Sec:Intro}
Nanoscale thermal transport is of considerable importance in the operation of modern electronic, optoelectronic, and thermoelectric devices.\cite{PopNanoRes2010,Cahill2014} In superlattices (SLs), multiple interfaces between different materials play a critical role in thermal transport.\cite{Cahill2003,Cahill2014} Advanced experimental techniques have enabled the measurements of the in-plane\cite{Hatta1985} and cross-plane\cite{Cahill1987,Panzer2009} thermal conductivity is SLs. Experiments show that the thermal conductivity of SLs is anisotropic and considerably lower than that of the constituent bulk materials.\cite{Yao1987,Chen1994,Yu1995,Capinski1996,Lee1997,Capinski1999,Huxtable2002,Vitiello2008,Luckyanova2013,Sood2014} Theoretical studies find that diffuse interface scattering is responsible for lowering of the in-plane (and, in part, the cross-plane) thermal conductivity, while the thermal boundary resistance (TBR) between adjacent layers is a key factor in the cross-plane thermal-conductivity reduction.\cite{Chen1997,Chen1998,Cahill2003,Cahill2014}

Superlattices based on III-V compound semiconductors have widespread use in optoelectronics. \cite{Faist1994,Kohler2002} In quantum cascade lasers (QCLs), self-heating is the main issue limiting the development of room-temperature (RT) continuous-wave lasing, which is exacerbated by the poor thermal conduction through hundreds of interfaces in a typical structure. \cite{Yao2012,Evans2008,Vitiello2008} A good understanding of the influence of interfaces on the thermal conductivity tensor of III-V SLs would enable advances in the design and modeling and optoelectronic devices for enhanced reliability.

The interfacial transport behavior is largely dependent on the material system and interface quality.\cite{Cahill2014} The acoustic mismatch model (AMM) and the diffuse mismatch model (DMM) have been traditionally used to calculate the phonon transmission coefficient and the resulting TBR of an interface.\cite{Khalatnikov1952,Swartz1989} These two models yeild the lower and upper limits of the TBR, respectively, but do not satisfactorily explain realistic experimental results.\cite{Cahill2003} Molecular dynamics  simulations \cite{Schelling2002,Stevens2007,Landry2009,Termentzidis2010,Chalopin2012,Liang2014} have provided valuable insights into heat transport across a number of solid-solid interfaces. The non-equilibrium Green's function technique (NEGF) has also been applied to describe the phonon dynamics,\cite{Zhang2007,Hopkins2009} generally without phonon-phonon scattering. In general, atomistic simulations are limited by computation burden, which makes it hard to study complicated SL structures, such as the active region of solid-state lasers.\cite{Faist1994,Kohler2002}

In this paper, we present a semiclassical model describing the full thermal conductivity tensor of III-V compound SL structures, and apply it to III-arsenide systems. The phonon transport inside each layer is captured by solving the phonon Boltzmann transport equation (PBTE) in the relaxation-time approximation (RTA), with rates describing the common internal scattering mechanisms as well as the partially diffuse scattering from the interfaces.\cite{Aksamija2013} The in-plane thermal conductivity is obtained from the layers connected in parallel, while the cross-plane conductivity is calculated from the layers and TBRs in series. The TBR of each interfaces is calculated by interpolating between the AMM and DMM transmission coefficients at the interface. Both the partially diffuse interface scattering and the AMM-DMM interpolation are described with the aid of the same momentum-dependent specularity parameter, in which there is a single adjustable parameter -- an effective interface rms roughness. The model can effectively describe complicated systems with an arbitrary number of interfaces and random layer thicknesses. Despite the model simplicity, the calculation results agree well with experimental data from multiple studies by different groups.\cite{Yao1987,Chen1994,Yu1995,Capinski1996,Capinski1999,Luckyanova2013,Sood2014} The model is also quite general: it is applicable to SLs in other material systems with good-quality interfaces and semiclassical phonon transport.\cite{Ravichandran2014,Lee1997,Huxtable2002}

This paper is organized as follows. Section~\ref{Sec:SL} describes the SL thermal transport model in detail: the role of interface roughness on the baseline layer conductivity that affects both cross-plane and in-plane conduction, as well as the additional effect it has on cross-plane transport through the TBR.  In Sec.~\ref{Sec:result}, we illustrate the robustness of the transport model by comparing our calculation results with a number of experiments on GaAs/AlAs and InGaAs/InAlAs superlattice systems, and we also calculate the thermal conductivity tensor in a quantum cascade laser active region. We conclude with Sec.~\ref{Sec:conclusion}. This paper is accompanied by electronic Supplementary Materials.


\section{Thermal Conductivity of III-V Superlattices} \label{Sec:SL}

A semiconductor superlattice (SL) is a periodic structure, with each period consisting of two or more thin layers of different materials. III-V semiconductor SLs have been widely used in electronic and photonic devices.\cite{Levine1987,Faist1994,Nakamura1998,Kohler2002} Experimental results on several material systems show that thermal conductivity of a SL is substantially lower than that of a weighted average of the constituent bulk materials.\cite{Yao1987,Chen1994,Yu1995,Capinski1996,Lee1997,Capinski1999,Huxtable2002,Luckyanova2013,Sood2014} Thermal transport in SLs also exhibits pronounced anisotropy: the cross-plane thermal conductivity (the thermal conductivity in the SL growth direction, normal to each planar layer) is much lower than the in-plane thermal conductivity.\cite{Cahill2014} Theoretical studies show that the interfaces between adjacent layers are responsible for both the overall reduction and the anisotropy of thermal conductivity. \cite{Chen1997,Chen1998,Schelling2002} Here, we offer a model that quantitatively captures both effects of the interfaces.

\begin{figure}
\includegraphics[width = 3 in]{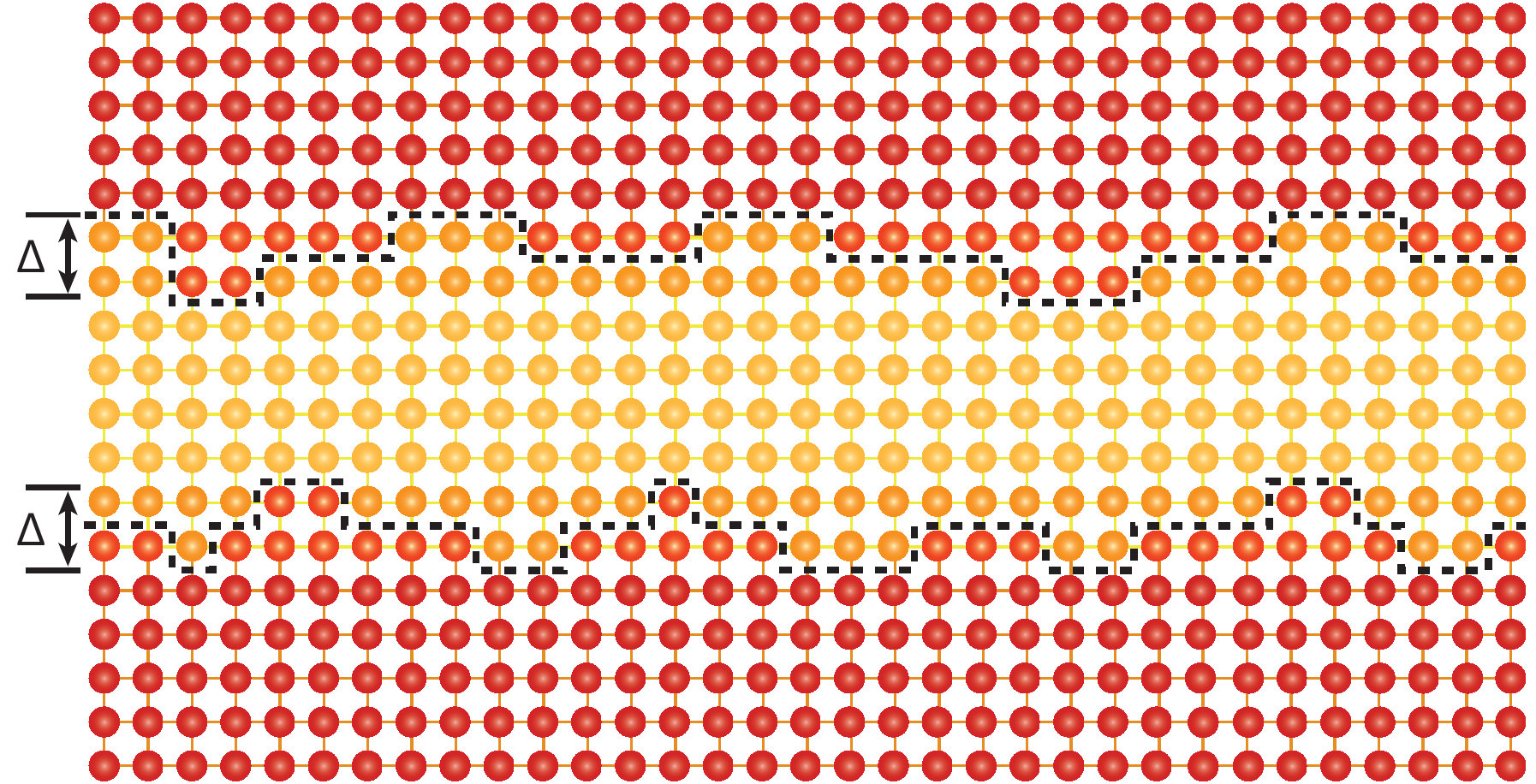}
	\caption{Even between lattice-matched crystalline materials, there exist nonuniform transition layers that behave as an effective atomic-scale interface roughness with some rms height $\Delta$. This effective interface roughness leads to phonon-momentum randomization and to interface resistance in cross-plane transport. }\label{fig:slsche}
\end{figure}

Even though a typical layer thickness in III-V SL structures is on the order of a few nanometers, we argue that coherent phonon transport can be neglected and that the semiclassical phonon Boltzmann equation provides an appropriate framework for analyzing heat flow in these systems over a range of temperatures. The reasons for this assertion are the following:

\begin{enumerate}
\item We are interested in the thermal conductivity of SLs near room temperature, where the phonon-phonon interaction is strong and breaks the phonon wave coherence.\cite{Ravichandran2014} The phonon coherence length in bulk GaAs at room temperature has been estimated to be smaller than 2 nm\cite{Chen1997,Chen1998} and this value will be even lower in ternary compounds owing to alloy scattering. Several SL structures we consider here  \cite{Yao1987,Chen1994,Yu1995,Capinski1996,Capinski1999,Luckyanova2013,Sood2014} have layers of thickness greater than or comparable to the phase-breaking length in individual layers, which implies that transport in them is largely incoherent. Indeed, experiments find that coherent transport features are important in GaAs/AlAs SLs below T$<$100 K.\cite{Luckyanova2012} In SLs with ternary III-V compounds, coherent transport phenomena would be important at even lower temperatures.

\item Even in best-quality lattice-matched SLs, there exists atomic scale interface roughness (Fig. \ref{fig:slsche})\cite{Jusserand1990,Ruf1994,Belk1997,Pillai2000,Robb2008,Luckyanova2012} which may break phonon phase coherence.\cite{Yang2003} Based on molecular dynamics, Wang \textit{et al.} \cite{WangAPL2015} showed that the thermal conductivity of SLs with rough interfaces increases monotonically with period length, in contrast to perfect SLs that feature nonmonotonic dependence. This finding confirms that phonons suffer from phase-breaking scattering in rough-interface SLs.

\item In QCLs, the SL has multiple periods, often called stages, with many layers in each stage. The layer thickness in each stage is highly variable, depending on the desired optoelectronic properties.\cite{Evans2008,Botez2013} Consequently, the QCL SL structure behaves as a nearly random multilayer system, which breaks phonon phase coherence.\cite{Wang2014,WangAPL2015}
\end{enumerate}

As a result of all the reasoning above, we do not consider phonon coherent transport or phonon confinement to analyze thermal transport. We use bulk dispersions and the phonon Boltzmann transport equation in the SL thermal conductivity calculations.

\subsection{The Twofold Influence of Effective Interface Roughness}

As mentioned briefly above, there will inevitably exist a few transition layers between adjacent materials in a SL structure.\cite{Jusserand1990,Ruf1994,Belk1997,Pillai2000,Robb2008,Luckyanova2012} Figure~\ref{fig:slsche} shows a schematic of interfaces between lattice-matched crystalline  layers in SLs. In the transition region, if we drew a line that separated the atoms of one crystal from those of the other, we would get a jagged boundary. Therefore, we model the interface with an effective interface rms roughness $\Delta$, which captures the basic properties of interfacial mixing. The thicker the transition layer, the higher the $\Delta$. Most III-V SLs are grown by molecular beam epitaxy (MBE)\cite{Cheng1997} or metal-organic chemical vapor deposition (MOCVD),\cite{Goetz1983} both well-controlled growth environments. As a result, all the interfaces in the SL should be nearly identical. Therefore, we use a single roughness $\Delta$ to model all the interfaces.

The probability of a phonon reflecting specularly from a rough interface is represented by a wave-number-dependent specularity parameter\cite{Soffer1967}
\begin{equation}\label{equ:spec}
p_\mathrm{spec}(\vec{q})=\exp(-4\Delta^2|\vec{q}|^2\cos^2\theta),
\end{equation}
where $|\vec{q}|$ is the magnitude of the wave vector, and $\theta$ represents the angle between $\vec{q}$ and the normal direction to the interface. This expression is nominally derived in the limit of uncorrelated roughness;\cite{Ziman1960,Soffer1967} but considering that more correlated surfaces scatter phonons more specularly,\cite{Maurer2015} surface correlation can effectively be captured by using a lowered $\Delta$.\cite{MaznevPRB2015}

Diffuse interface scattering affects all phonons in the SL and influences phonon mode occupation inside each layer. \cite{Aksamija2013} The effect on interface roughness on mode population can be captured by solving the PBTE with appropriate boundary conditions. The result is an effective interface scattering rate \cite{Aksamija2013} that captures the interplay between internal mechanisms (normal and umklapp three-phonon, isotope, alloy, dopant, and electron scattering; see Supplementary Materials) and interface roughness in a layer of thickness $L$:

\begin{subequations}
	\begin{equation}\label{eq:interface rate}
\tau_\mathrm{interface}^{-1}(\vec{q})=\frac{v_\mathrm{b,\perp}(\vec{q})}{L}\frac{F_\mathrm{p}(\vec{q},L)}{1-\frac{\tau_\mathrm{b,internal}(\vec{q})v_\mathrm{b,\perp}(\vec{q})}{L}F_\mathrm{p}(\vec{q},L)},
	\end{equation}
	where
	\begin{equation}
F_\mathrm{p}(\vec{q},L)=\frac{[1-p_\mathrm{spec}(\vec{q})]\{1-\exp[-L/\tau_\mathrm{b,internal}(\vec{q})v_\mathrm{b,\perp}]\}}{1-p_\mathrm{spec}(\vec{q})\exp[-L/\tau_\mathrm{b,internal}(\vec{q})v_\mathrm{b,\perp}]}
	\end{equation}
\end{subequations}
is a mode-dependent scaling factor. Here, b denotes the phonon branch and $\vec{q}$ its wave vector, $v_\mathrm{b,\perp}$ is the component of the phonon group velocity normal to the interface, and $\tau_\mathrm{b,internal}(\vec{q})$ is the total relaxation time due to internal scattering mechanisms in the layer (see Supplementary Materials). It is noteworthy that the effective rate of interface scattering (\ref{eq:interface rate}) depends on both roughness and the relative size of the layer thickness ($L$) to the mean free path for internal scattering ($\tau_\mathrm{b,internal}(\vec{q})v_\mathrm{b,\perp}$): for very thin layers ($L/\tau_\mathrm{b,internal}(\vec{q})v_\mathrm{b,\perp}<<1$) the phonon ``sees'' both interfaces of a layer ($\tau_\mathrm{interface}^{-1}(\vec{q})\rightarrow 2\frac{v_\mathrm{b,\perp}(\vec{q})}{L}\frac{1-p_{\mathrm{spec}(\vec{q})}}{1+p_{\mathrm{spec}(\vec{q})}}$, a well-known expression derived by Ziman\cite{Ziman1960}), while for very thick layers ($L/\tau_\mathrm{b,internal}(\vec{q})v_\mathrm{b,\perp}>>1$), the phonon will scatter many times due to internal mechanisms between successive interactions with interfaces, as if the interfaces were completely independent ($\tau_\mathrm{interface}^{-1}(\vec{q})\rightarrow \frac{v_\mathrm{b,\perp}(\vec{q})}{L}[1-p_{\mathrm{spec}(\vec{q})}]$). For details, see Ref. [\onlinecite{Aksamija2013}]. Through this additional effective scattering rate, rough interfaces that bound each layer affect phonon population and thus influence both in-plane and cross-plane thermal transport.\cite{Aksamija2013} This is the first aspect of interfacial influence on thermal transport in SLs.

The cross-plane thermal conductivity bears an additional influence of the interfaces.\cite{Swartz1989,Simkin2000} In order to carry heat along the cross-plane direction, phonons must cross interfaces. As there are two different materials on the two sides of the interface, the phonon transmission probability through the interface is not unity, and a thermal boundary resistance emerges. There have been two widely accepted models  -- the acoustic mismatch model (AMM)\cite{Khalatnikov1952,Little1959} and the diffuse mismatch model (DMM) -- for the calculation of the phonon transmission coefficient and the TBR.\cite{Swartz1989}

From the AMM point of view, the interface between two isotropic media is treated as a perfect plane and the phonons are treated as plane waves. The AMM transmission coefficient is the ratio of transmitted to injected heat flux, and is calculated upon solving the elastic continuum equation with appropriate boundary conditions (continuity of the normal component of the wave number, which will yield a Snell's law analogue, and continuity of  the velocity field and tangential force). The AMM transmission coefficient for a phonon going from material 1 to material 2 can be expressed as:
\begin{equation}
t_\mathrm{b,1\rightarrow 2}^\mathrm{AMM}(\vec{q})=\frac{4Z_\mathrm{b,1}^\perp(\vec{q})Z_\mathrm{b,2}^\perp(\vec{q})}{\left[Z_\mathrm{b,1}^\perp(\vec{q}) +Z_\mathrm{b,2}^\perp(\vec{q})\right]^2},
\end{equation}
where $Z_\mathrm{b,1/2}^\perp=\rho_{1/2} v_\mathrm{b,1/2}^\perp(\vec{q})$  are the perpendicular acoustic impedances of sides 1 and 2. $\rho$ is the mass density of a material. Here, we work with full phonon dispersions, so it is hard to achieve detailed balance, i.e., conserve both momentum and energy for a phonon going through an interface. However, the lattice structures and dispersion curves for III-As are very similar, so we simply momentum and the resulting error in energy conservation is quite small.

On the other hand, in the DMM, the assumption is that the coherence is completely destroyed at the interface: a phonon loses all memory about its velocity and randomly scatters into another phonon with the same energy. The transmission coefficient can be derived from the principle of detailed balance as\cite{Reddy2005}
\begin{equation}
t_{1\rightarrow 2}^\mathrm{DMM}(\vec{q})=\frac{v_\mathrm{b,2}(\vec{q})D_2(\omega_1(\vec{q}))}{v_\mathrm{b,2}(\vec{q})D_2(\omega_1(\vec{q}))+v_\mathrm{b,1}(\vec{q})D_1(\omega_1(\vec{q}))},
\end{equation}
where $D_1(\omega)$ and $D_2(\omega)$ are the phonon densities of states in materials 1 and 2, respectively.

In reality, for a high-quality interface like that in a III-V SL structure, phonon interface scattering is neither purely specular nor completely diffuse; consequently, the AMM overestimates while the DMM underestimates the transmission coefficient. \cite{Koh2009} In order to accurately model the TBR in a large temperature range and for various interfaces, we will \textit{interpolate} between the two models for the transmission coefficient.\cite{Chen1998,Kazan2011} \emph{We posit that the specularity parameter (\ref{equ:spec}) can also be used to give weight to the probability of phonon transmission without momentum randomization, i.e., to the AMM transmission coefficient.} In other words, we introduce an effective phonon transmission coefficient as
\begin{equation}\label{equ:transmission}
t_{\mathrm{b}}(\vec{q})=p_\mathrm{spec}(\vec{q})\cdot t_{\mathrm{b}}^\mathrm{AMM}(\vec{q})+\left[1-p_\mathrm{spec}(\vec{q})\right]\cdot t_{\mathrm{b}}^\mathrm{DMM}(\vec{q}).
\end{equation}
This coefficient captures both the acoustic mismatch and the momentum randomization at a rough interface between two media. The rougher the interface, the lower the specularity parameter, and therefore the higher the TBR. The TBR will only be picked up by the phonons trying to cross an interface, thus having an influence on cross-plane transport only. This is the second effect the roughness has on the thermal transport.

We note that the above discussion holds for acoustic phonons, which are the dominant carriers of heat in semiconductors. The role of optical phonons in bulk heat transport has recently been highlighted,\cite{Lindsay2013} but they are relatively minor contributors to bulk heat transport owing to the low occupation number and group velocity. It is also unclear how optical phonons behave when crossing boundaries, but it is likely that their transmission is highly suppressed because their existence hinges of good crystallinity. A recent paper by Ong and Zhang supports this assertion. \cite{OngPRB2015} Therefore, optical phonons are neglected in this study.

\subsection{Calculation of In-plane and Cross-plane Thermal Conductivities} \label{sub:incross}

First, each layer's thermal conductivity is calculated in the same way as the bulk thermal conductivity of a material (see Sec. S-I in Supplementary Materials), but with an additional scattering rate (\ref{eq:interface rate}) due to the presence of interfaces.\cite{Aksamija2013} The layer thermal conductivity obtained this way will already be lower than the bulk thermal conductivity of the same material.

Second, the TBR is calculated using a transmission coefficient interpolated from the AMM and the DMM values. The TBR from material 1 to material 2, denoted $R_{1\rightarrow 2}$, is given by
\begin{equation}\label{equ:conductance}
R_{1\rightarrow 2}^{-1}=\frac{1}{2}\sum\limits_{b,\vec{q}}\frac{v_\mathrm{b,1,\perp}(\vec{q})C_\mathrm{b,T}(\vec{q})t_{1\rightarrow 2}(\omega_1(\vec{q}))}{1-\frac{1}{2}\left<t_{1\rightarrow 2}(\omega_1(\vec{q}))+t_{2\rightarrow 1}(\omega_1(\vec{q}))\right>}.
\end{equation}
The denominator in the expression is a correction factor introduced following the modified definition of temperature of Simons\cite{Simons1974} and Zeng and Chen,\cite{Zeng2003} as the phonon distribution at the interface is far from equilibrium. The correction ensures that the TBR vanishes at a fictitious interface inside a material. Here $\left<t_{1\rightarrow 2}(\omega_1(\vec{q}))+t_{2\rightarrow 1}(\omega_1(\vec{q}))\right>$ represents the average value of transmission coefficients over the Brillouin zone.

With properly calculated layer thermal conductivity and the TBR, the in-plane and cross-plane thermal conductivity of a SL with two layers per period can be written as\cite{Simkin2000,Aksamija2013}
\begin{subequations}
	\begin{align}
	\kappa_\mathrm{in-plane} &=\frac{L_1\kappa_1+L_2\kappa_2}{L_1+L_2},\\
	\kappa_\mathrm{cross-plane} &= \frac{L_1+L_2}{\frac{L_1}{\kappa_1+}+\frac{L_2}{\kappa_2}+(R_{1\rightarrow 2}+R_{2\rightarrow 1})},
	\end{align}
where $L_1$ and $L_2$ are the layer thicknesses of materials 1 and 2, respectively, while $\kappa_1$ and $\kappa_2$ are the corresponding layer thermal conductivities. $R_{1\rightarrow 2}$ and $R_{2\rightarrow 1}$ represent the TBRs from layer 1 to layer 2 and from layer 2 to layer 1. The expressions can be extended to the situation of a SL with $n$ layers of thicknesses $L_{i}$ ($i=1,\ldots,n$):
\begin{align}
\kappa_\mathrm{in-plane} &=\frac{\sum_{i=1}^{n}L_{i}\kappa_{i}}{\sum_{i=1}^{n}L_{i}},\\
\kappa_\mathrm{cross-plane} &= \frac{\sum_{i=1}^{n}L_{i}}{\sum_{i=1}^{n} L_{i}/\kappa_{i}+R_{i\rightarrow i+1}},\label{eq:cross kappa}
\end{align}
\end{subequations}
with the understanding that $R_{n\rightarrow n+1}\equiv R_{n\rightarrow 1}$, owing to periodic boundary conditions (i.e., after the last layer $n$ comes layer $1$ again). Considering that the TBRs are generally not symmetric ($R_{i\rightarrow j}\neq R_{j\rightarrow i}$), the cross-plane thermal conductivity is not the same in both directions, so SLs can exhibit thermal rectification properties.

\section{Results and Comparison with Experiments} \label{Sec:result}

\subsection{GaAs/AlAs Superlattices}\label{sub:GaAs/AlAs}

We have compared the results from our simple model with several experimental results by different groups on both in-plane\cite{Yao1987,Yu1995,Luckyanova2013} and cross-plane\cite{Chen1994,Capinski1996,Capinski1999,Luckyanova2013,Sood2014} thermal conductivity of III-arsenide SLs and obtained good agreement.

\begin{figure}
\includegraphics[width = 3 in]{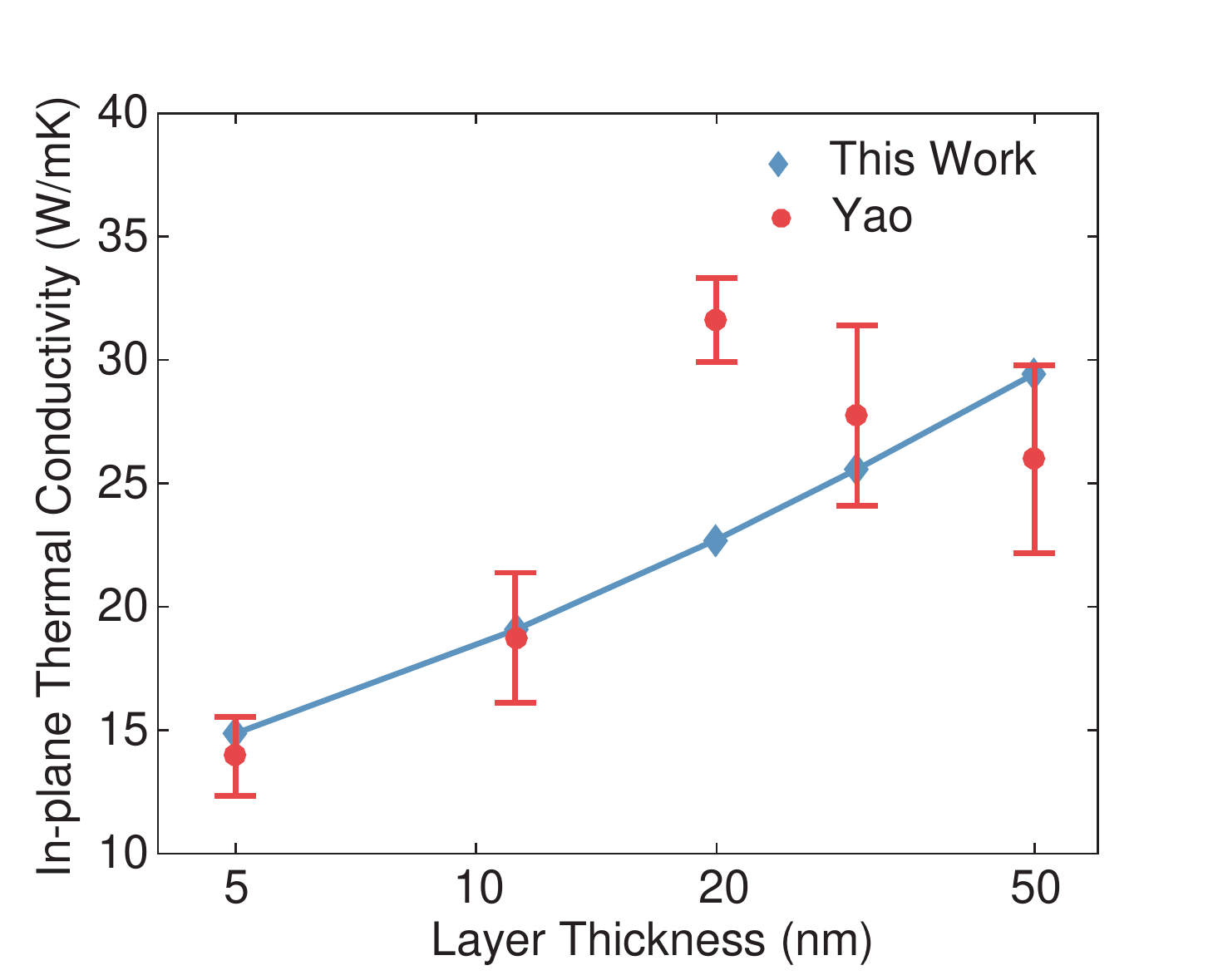}
	\caption{In-plane thermal conductivity of GaAs/AlAs superlattices as a function of layer thickness. Red dots are the experimental data from Yao \cite{Yao1987} and blue diamonds are our calculated data with $\Delta=6\,\mathrm{\AA}$.}
	\label{fig:yao}
\end{figure}

Figure~\ref{fig:yao} shows the RT in-plane thermal conductivity of GaAs/AlAs SLs with various layer thicknesses. To compare with Yao's data,\cite{Yao1987} we set the effective interface roughness to 6 $\mathrm{\AA}$.
The in-plane thermal conductivity should first increase monotonically with increasing layer thickness, then saturate at the average bulk value of roughly 66 W/mK. The measurement is non-monotonic and appears to saturate at a lower value, which  Yao \cite{Yao1987} suggested stems from pronounced interfacial mixing and thus considerable alloy scattering of phonons between layers. We note that our model does not capture significant interfacial mixing and is instead  suitable for high quality interfaces with only atomic-scale roughness.

\begin{figure}
\includegraphics[width = 3 in]{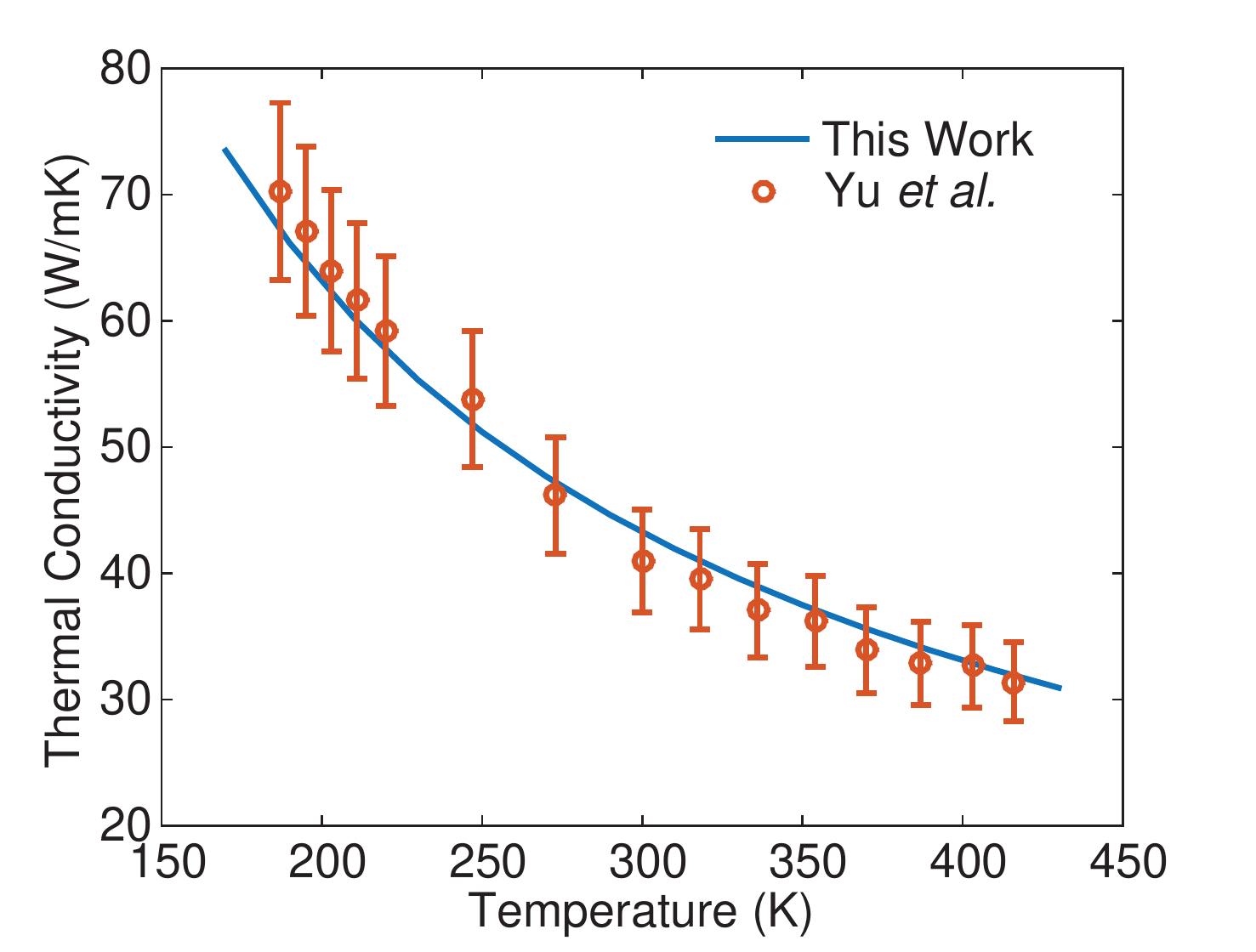}
	\caption{In-plane thermal conductivity of a GaAs/AlAs superlattice (layer thickness 70 nm) as a function of temperature. Red symbols show the experimental results of Yu \textit{et al.}\cite{Yu1995} and the blue curve shows the calculations from our model with $\Delta=3.7~\mathrm{\AA}$.}
	\label{fig:yu}
\end{figure}

Figure~\ref{fig:yu} shows the in-plane thermal conductivity of a GaAs/AlAs SL with a layer thickness of 70 nm at various temperatures. The symbols are the experimental results reported by Yu \textit{et al.}\cite{Yu1995} and the line is our calculation with $\Delta=3.7~\mathrm{\AA}$. The calculation agrees well with experiment over a wide temperature range.

\begin{figure}
\includegraphics[width = \columnwidth]{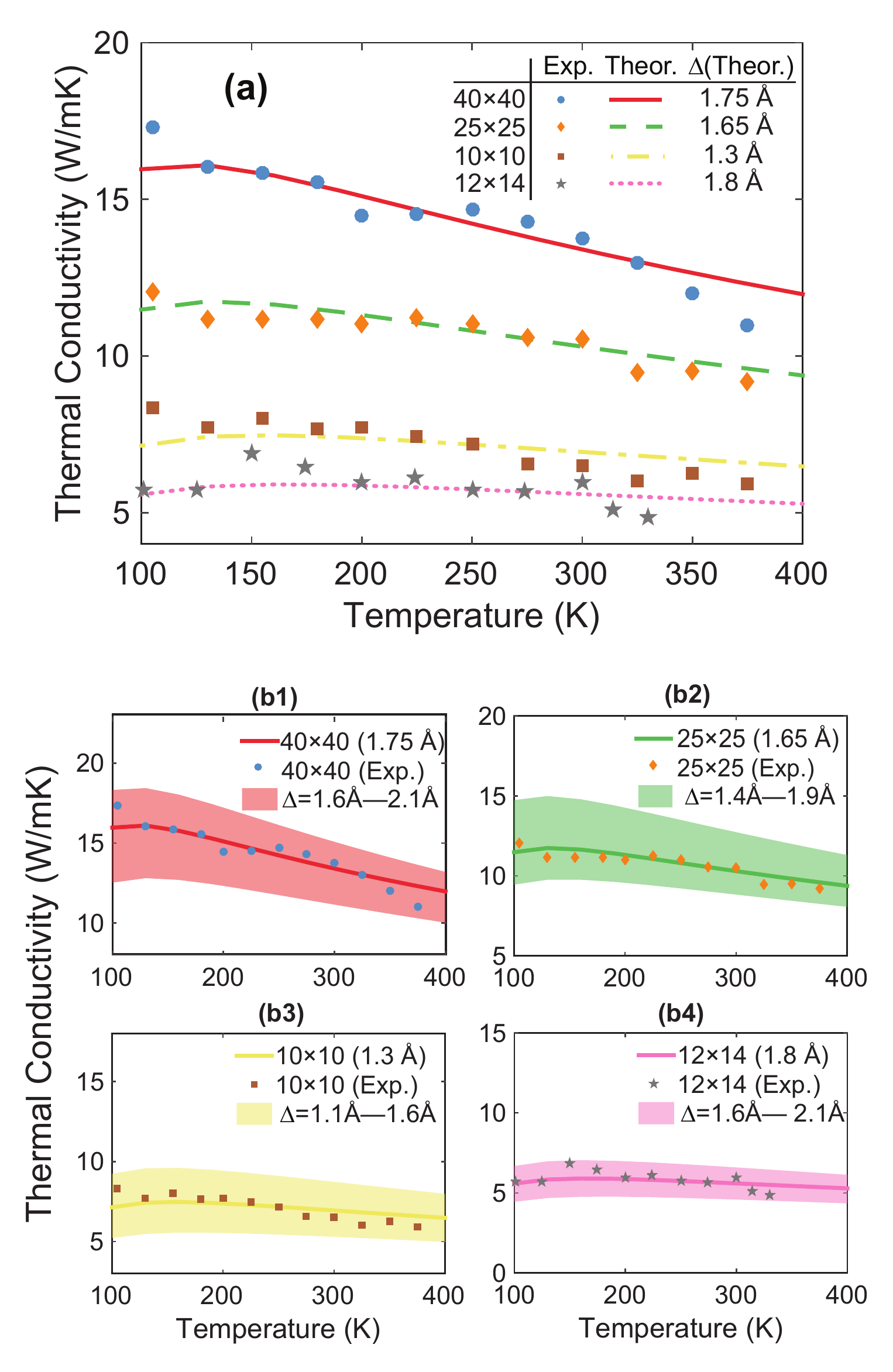}
	\caption{(a) Cross-plane thermal conductivity of GaAs/AlAs superlattices as a function of temperature. Blue circles, orange diamonds, and brown squares show the measured cross-plane thermal conductivity data for 40$\times$40, 25$\times$25, and 10$\times$10 SLs from Capinski \textit{et al.}\cite{Capinski1999} Grey stars are the cross-plane thermal conductivity data for a 12$\times$14 SL from Capinski and Maris.\cite{Capinski1996} The corresponding curves are calculated based on our model, with the optimal effective rms roughness $\Delta$ denoted in the legend. (b1)--(b4) Sensitivity of the cross-plane thermal conductivity to interface roughness. Each panel represents one set of data from (a), along with the optimal fit based on our model [also given in (a)], and a shaded area depicting the range of thermal conductivities that would be obtained by varying rms roughness $\Delta$ by 0.5 $\mathrm{\AA}$. }
	\label{fig:capinski}
\end{figure}

Figure~\ref{fig:capinski} shows the cross-plane thermal conductivity of GaAs/AlAs SLs with various layer thicknesses and from 100 K to 400 K. Symbols show the experimental results reported by Capinski and Maris\cite{Capinski1996} and Capinski \textit{et al.}\cite{Capinski1999} The corresponding curves are obtained from our model. We first set the layer thicknesses to those reported in experiments and then vary the effective roughness to get the best fit; panels (b1)--(b4) in Fig.~\ref{fig:capinski} illustrate how sensitive thermal conductivity is to rms-roughness variation. The optimal-fit rms roughness is 1.75 $\mathrm{\AA}$, 1.65 $\mathrm{\AA}$, 1.3 $\mathrm{\AA}$, and 1.8 $\mathrm{\AA}$ for the 40$\times$40, 25$\times$25, 10$\times$10, and 12$\times$14 SLs, respectively [$\times$ is the notation in these two experimental papers]. The small values of the rms roughness are in keeping with high-quality interfaces, featuring large-area atomically flat terraces. The cross-plane thermal conductivity varies very little as the temperature changes, indicating that the TBR indeed dominates thermal transport across layers.

Luckyanova \textit{et al.}\cite{Luckyanova2013} recently measured both the in-plane and the cross-plane thermal conductivity of a GaAs/AlAs SL. Our calculation for the same structure and the experimental results are shown in Table~\ref{tab:luck}. All the calculation results used an effective interface rms roughness of 1.1 {\AA} for the 2-nm system and 1.9 {\AA} for the 8-nm one, which results in good agreement for the cross-plane conductivity; however, the measured in-plane thermal conductivity is considerably lower than the calculation. In fact, the experimental data from Luckyanova \textit{et al.}\cite{Luckyanova2013} shows a great discrepancy with all the previous experiments on similar systems.\cite{Yao1987,Yu1995,Capinski1996,Capinski1999} For example, the in-plane thermal conductivity of their 8-nm SL is considerably smaller than that of the 5-nm SL in Yao's paper,\cite{Yao1987} which is counterintuitive and does not agree with well-established trends of increasing thermal conductivity with increasing layer thickness. Furthermore, the cross-plane thermal conductivity (8.7$\pm0.4$ W/mK) is considerably smaller than that of Capinski \textit{et al.} (10.52 W/mK) with similar layer thickness. The earlier experiments\cite{Yao1987,Capinski1999} should have worse or at best equivalent interface quality to the samples in the most recent work,\cite{Luckyanova2013} owing to the development in growth techniques that happened over the past few decades; yet, older samples show higher conductivities. Luckyanova \textit{et al.}\cite{Luckyanova2013} also performed density functional perturbation theory (DFPT) simulation, the results of which are about twice what they measured.
\begin{table}
	\begin{tabular}{ c@{\hspace{1.0em}}c@{\hspace{0.7em}}c@{\hspace{1.0em}}c@{\hspace{0.7em}}c }
		\hline\hline
		layer thickness & \multicolumn{2}{c}{2 nm} & \multicolumn{2}{c}{8 nm} \\
		& exp & cal & exp & cal\\
		\hline
		$\kappa_\mathrm{in-plane}$ & $8.05\pm0.48$ & 25.03 & $11.4\pm0.46$ & 22.78\\
		$\kappa_\mathrm{cross-plane}$ & $6.5\pm0.5$ & 6.38 & $8.7\pm0.4$ & 8.59 \\
		\hline\hline
	\end{tabular}
	\caption{Comparison of experimental results from Luckyanova \textit{et al.}\cite{Luckyanova2013} and our calculated data for GaAs/AlAs SLs with layer thickness of 2 nm and 8 nm. In the calculation, we assume an interface rms roughness of 1.1 {\AA} for the 2-nm system and 1.9 {\AA} for the 8-nm one.}
	\label{tab:luck}
\end{table}

\subsection{InGaAs/InAlAs Superlattices}\label{sub:InGaAs/InAlAs}

\begin{figure}
\includegraphics[width = 3 in]{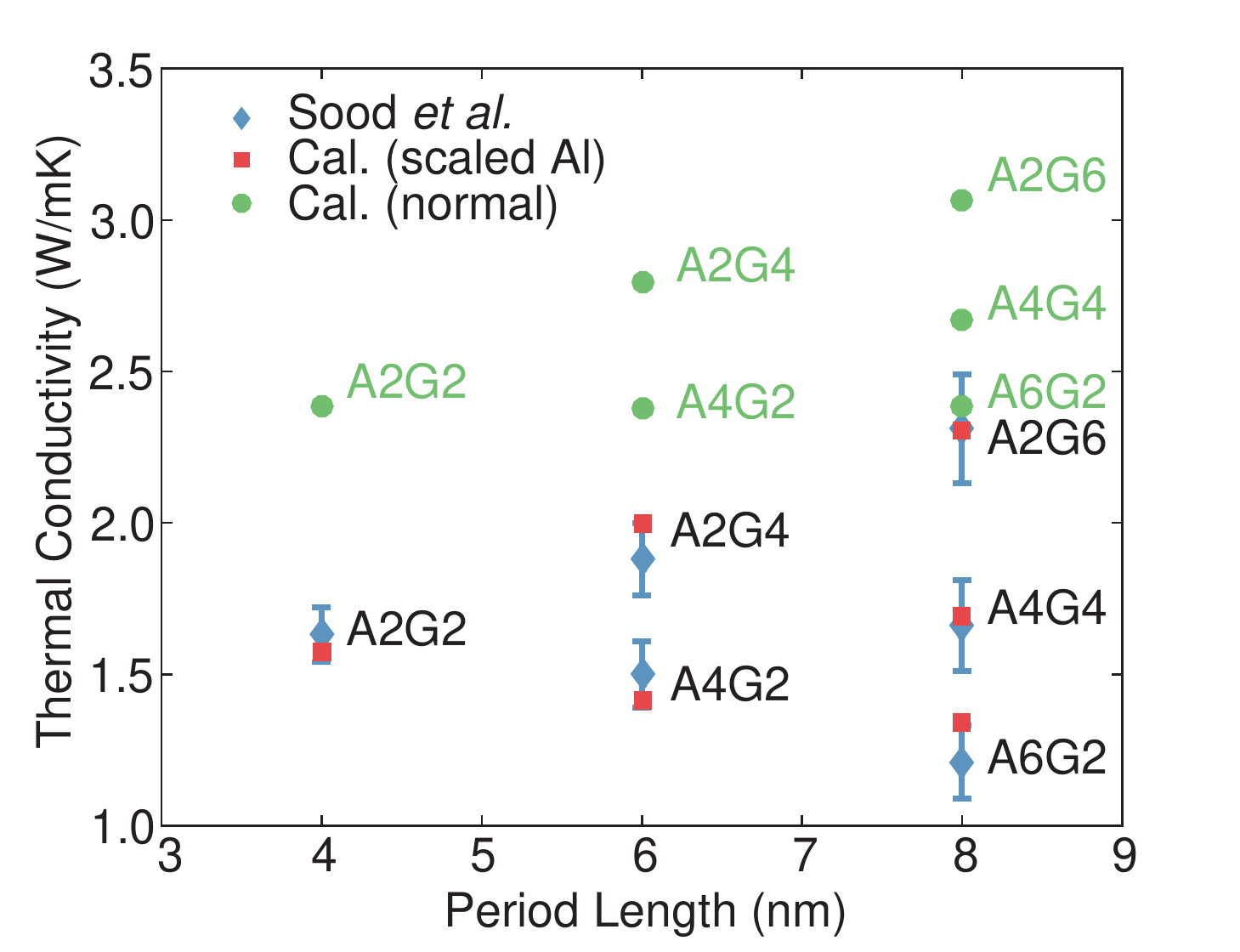}
	\caption{Cross-plane thermal conductivity of $\mathrm{In}_{0.53}\mathrm{Ga}_{0.47}\mathrm{As}/\mathrm{In}_{0.52}\mathrm{Al}_{0.48}\mathrm{As}$ SLs as a function of the period length. The notation AmGn represents a SL structure with the InAlAs and InGaAs layer thicknesses of m and n nanometers, respectively. Blue diamonds show the experimental data from Sood \textit{et al.}\cite{Sood2014}, green dots show our calculation with regular scattering rates, and red squares show the calculation results with artificially increased bulk rates for InAlAs. In both calculations, $\Delta$ is chosen to be $0.5~\mathrm{\AA}$, in keeping with the perfect interface quality revealed in X-ray diffraction experiments.}
	\label{fig:sood}
\end{figure}

Sood \textit{et al.}\cite{Sood2014} studied the RT cross-plane thermal conductivity of lattice-matched InGaAs/InAlAs SLs (In$_{0.53}$Ga$_{0.47}$As/In$_{0.52}$Al$_{0.48}$As) with varying layer thicknesses. They used the notation AmGn to represent a SL structure with the InAlAs and InGaAs layer thicknesses of n and m nanometers, respectively. Six different SL structures (A2G2, A2G4, A2G6, A4G2, A4G4, A6G2) were measured and these experimental results are reproduced as blue diamonds in Figure~\ref{fig:sood}.

We show our calculation results in Figure~\ref{fig:sood}. The green circles are the results with our calculated bulk rates (see Supplementary Materials). We assume very small roughness $\Delta=0.5~\mathrm{\AA}$,  in keeping with the X-ray diffraction measurements that show nearly perfect interface quality. We note that the green data points are higher than the measurement, but that the trend with the period length is the same as in experiment. Indeed, increasing the interface roughness would significantly and adversely affect the slope of the thermal conductivity with increasing period length. Therefore, we assert that the reason for the discrepancy has to do with internal scattering in InAlAs.

Namely, from their data, Sood \textit{et al}.\cite{Sood2014} extract the bulk thermal conductivities of InGaAs and InAlAs to be 5 W/mK and 1 W/mK, respectively. While our calculated bulk thermal conductivity of InGaAs (5.17  W/mK) matches experiment, we calculate the bulk conductivity of InAlAs to be  3.1 W/mK, considerably higher than what Sood \textit{et al.} reported. Unfortunately, there is no direct experimental measurement of the thermal conductivity of InAlAs.

In red squares, we artificially increase the internal scattering rate of InAlAs so that its bulk thermal conductivity is around 1 W/mK, in keeping with Sood \textit{et al}.\cite{Sood2014}, and we keep the interface scattering rate as before, corresponding to very small $\Delta=0.5~\mathrm{\AA}$ for good-quality interfaces. We see that the red squares agree very well with experimental data, both quantitatively and in the trend with with increasing period length. Considering that the normally calculated thermal conductivity for InAs and AlAs agree with experiment, and that our calculation for InGaAs agrees well with other measurements as well with the value extracted by Sood \textit{et al}.\cite{Sood2014} we believe there is a nontrivial aspect of alloy scattering in InAlAs that leads to much lower bulk conductivity of InAlAs than anticipated. Namely, the standard mass-difference scattering model based on the work of Abeles\cite{Abeles1963} and Adachi\cite{Adachi1983} (see Supplementary Materials for details) is rooted in the perturbation theory. In InAlAs with nearly equal amounts of InAs and AlAs, the difference between the cation masses exceeds the average cation mass owing to the large atomic-mass difference between In and Al, which we believe makes a perturbative approach invalid. This is not a problem in either AlGaAs or InGaAs, where the cation mass disparity is not as dramatic as in InAlAs and the perturbative mass-difference calculation agrees well with measurements (see Supplementary Materials). The hypothesis that the perturbative mass-difference approach fails in InAlAs would have to be tested in atomistic simulations, which are beyond the scope of this work, and in direct experimental measurements of the bulk thermal conductivity of InAlAs.

\subsection{Application to Thermal Modeling of a Quantum Cascade Laser} \label{sub:qcl}
The quantum cascade laser is a common application of III-V SLs. The active region of a QCL consists of tens of repeated stages, where each stage consists of tens of thin layers.\cite{Faist1994} Thermal modeling of such devices has always been challenging because of the great anisotropy in thermal transport caused by the SL structure.\cite{Cahill2014,Vitiello2008} It is difficult to accurately describe the in-plane and cross-plane thermal conductivity of such structures with existing simulation methods because of the complicated layer structure inside one stage.

\begin{figure}
\includegraphics[width = 3 in]{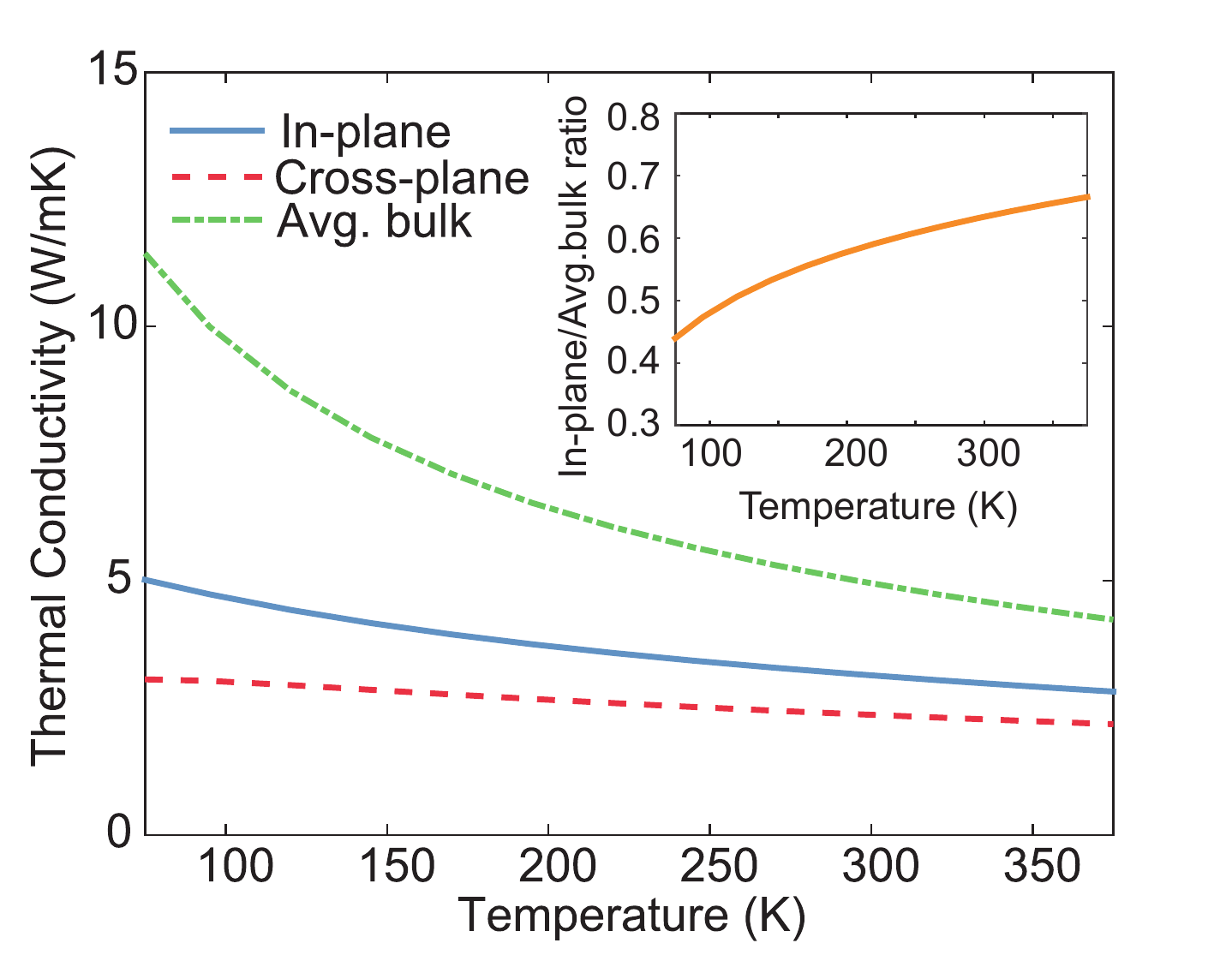}
	\caption{Thermal conductivity of a typical QCL active region\cite{Lops2006} as a function of temperature. A single stage consists of 16 alternating layers of $\mathrm{In}_{0.53}\mathrm{Ga}_{0.47}\mathrm{As}$ and $\mathrm{In}_{0.52}\mathrm{Al}_{0.48}\mathrm{As}$. Blue solid curve, red dashed curve, and green dash-dot curve are showing the calculated in-plane, cross-plane, and the averaged bulk thermal conductivity, respectively. $\Delta=1~\mathrm{\AA}$ in the calculations. The inset shows the ratio between the calculated in-plane and the averaged bulk thermal conductivities.}
	\label{fig:lops}
\end{figure}

It is often assumed that the in-plane thermal conductivity of a SL structure is 75\% of the corresponding bulk average in all temperature ranges. Under this assumption, a constant cross-plane thermal conductivity is used as a tunable parameter to fit the measured temperature profile.\cite{Lops2006,Evans2008,Lee2010} We show below that the assumption about in-plane thermal conductivity being 75\% of the weighted bulk value does not generally hold. This ratio is lower and temperature dependent, varying from 40\% to 70\% as the temperature raises from 100 K to 400 K (inset of Fig.~\ref{fig:lops}).  Overestimating the in-plane leads to somewhat underestimating the cross-plane thermal conductivity based on a fit to a temperature profile.\cite{Lops2006,Evans2008,Lee2010}

Figure~\ref{fig:lops} shows our calculated in-plane and cross-plane thermal conductivity for a typical QCL active region.\cite{Lops2006} A single stage of the SL structure consists of 16 alternating layers of $\mathrm{In}_{0.53}\mathrm{Ga}_{0.47}\mathrm{As}$ and $\mathrm{In}_{0.52}\mathrm{Al}_{0.48}\mathrm{As}$, and the interface roughness is set to 1 $\mathrm{\AA}$. The in-plane thermal conductivity is 65\% of the bulk value at RT, and the calculated cross-plane thermal conductivity is 2.37 W/mK, close to but slightly higher than the extracted experimental value of 2.2 W/mK. It is reasonable because their estimated in-plane thermal conductivity is slightly higher. Furthermore, the anisotropy of thermal conductivity is not overly pronounced: the ratio between the in- and cross- plane value is only about a factor of 2 in InGaAs/InAlAs-based QCLs here. The ratio is greater ($\sim$5.5 at 100 K) for GaAs/AlGaAs-based QCLs,\cite{Evans2008} because the in-plane thermal conductivity is much higher as GaAs is a binary material. The cross-plane thermal conductivity is fairly insensitive to temperature variation, which underscores the dominance of the temperature-insensitive TBR on cross-plane heat conduction. (We note that we have used the mass-difference alloy scattering model, which as discussed above (Sec. \ref{sub:InGaAs/InAlAs}) may underestimate alloy scattering in InAlAs.)

Because there is no universal relationship between the in-plane, cross-plane, and the weighted average of bulk thermal conductivity, it is difficult to get a sense of what the thermal conductivity would be for a new structure. In such cases, the model we presented here can provide a fairly quick calculation to help with thermal modeling of novel QCL structures.\cite{Shi2012}

\section{Conclusion} \label{Sec:conclusion}

By solving the phonon Boltzmann transport equation under the RTA, we analyzed thermal transport in III-As SL structures. The calculation of the thermal conductivity tensor in superlattices involves each layer's conductivity, itself affected by the impact of diffuse interface scattering on phonon populations, as well as explicit thermal boundary resistance that only affects cross-plane thermal transport. We calculate the TBR between interfaces based on interpolating the transmission coefficient between AMM and DMM, where the specularity parameter (traditionally used to describe diffuse scattering) is also used as the AMM weight in the interpolation. Therefore, with a single free parameter -- the effective interface rms roughness $\Delta$ (often ranging from 0.5 $\mathrm{\AA}$ to 6 $\mathrm{\AA}$ for high-quality III-As interfaces) -- we captured the transport properties of multiple GaAs/AlAs and InGaAs/InAlAs SL structures over a wide temperature range (70 K to 400 K). We have also applied the model to a typical QCL structure, in good agreement with experiment.

The presented model is fairly simple yet quite accurate, especially when used with full phonon dispersions. It can be very useful for thermal modeling complicated QCL structures, with many interfaces. The model is also applicable to other material systems where SLs have good-quality interfaces and phonon transport can be considered incoherent.

\begin{acknowledgments}
The authors thank L. Mawst and D. Botez for their comments on the manuscript,  and L. N. Maurer for his help with the revisions. This work was supported by the NSF under Award No. 1201311. The simulation work was performed using the compute resources and assistance of the UW-Madison Center For High Throughput Computing (CHTC) in the Department of Computer Sciences.
\end{acknowledgments}

%

\clearpage

\setcounter{secnumdepth}{3}

\thispagestyle{empty}
\setcounter{page}{1}
\setcounter{equation}{0}
\setcounter{figure}{0}
\setcounter{section}{0}
\setcounter{subsection}{0}
\setcounter{subsubsection}{0}
\setcounter{table}{0}

\renewcommand\thesection{SECTION S-\Roman{section}}

\renewcommand\thesubsection{S-\Roman{section}.\Alph{subsection}}
\renewcommand\thesubsubsection{S-\Roman{section}.\Alph{subsection}.\arabic{subsubsection}}

\renewcommand\thefigure{S\arabic{figure}}

\renewcommand\thetable{S-\Roman{table}}

\renewcommand\theequation{S\arabic{equation}}



\begin{widetext}
\noindent{\large\bf Supplementary Materials to \\
``Thermal conductivity of III-V semiconductor superlattices"}

\vspace{0.5 cm}
\noindent S. Mei and I. Knezevic\\
Department of Electrical and Computer Engineering, University of Wisconsin-Madison, Madison, WI 53706, USA\\
\end{widetext}


\section{Thermal Conductivity of III-V Binary and Ternary Compounds with Full Phonon Dispersion} \label{Sec:method}
\subsection{Thermal Conductivity Tensor} \label{sub:tctensor}
We are interested in thermal transport in III-V compound semiconductors, where acoustic phonons carry most of the heat.\cite{S-Morelli2002,S-Lindsay2013} The full thermal conductivity tensor of a crystalline semiconductor at temperature $T$ can therefore be calculated by summing over the contributions from all phonon wavevectors and branches,\cite{S-Klemens1958}
\begin{equation}\label{eq:bulktc}
\kappa^{\alpha\beta}(T)=\sum_{\mathrm{b},\vec{q}}C_\mathrm{b}(\vec{q},T)\tau_\mathrm{b}(\vec{q},T)v_\mathrm{b}^\alpha(\vec{q})v_\mathrm{b}^\beta(\vec{q}),
\end{equation}
where $C_\mathrm{b}(\vec{q},T)$ is the phonon heat capacity for mode b, given by
\begin{equation}\label{eq:heatcap}
C_\mathrm{b}(\vec{q},T)=\frac{\left[\hbar\omega_\mathrm{b}(\vec{q})\right]^2}{k_\mathrm{B}T^2}\frac{e^{\hbar\omega_\mathrm{b}(\vec{q})/k_\mathrm{B}T}}{\left[e^{\hbar\omega_\mathrm{b}(\vec{q})/k_\mathrm{B}T}-1\right]^2}.
\end{equation}
$\tau_\mathrm{b}(\vec{q},T)$ is the total phonon relaxation time and $v_\mathrm{b}^\alpha(\vec{q})$ is the mode- and wavevector-specific phonon group velocity along the $\alpha$ direction. Both the heat capacity and the group velocity are calculated from the exact phonon dispersion relation using the adiabatic bond charge model (ABCM).\cite{S-Weber1974,S-Rustagi1976,S-Tutuncu2000} We numerically evaluate Eq.~(\ref{eq:bulktc}) over the first Brillouin zone (1BZ) to obtain one entry in the bulk thermal conductivity tensor. When dealing with ternary compounds ($\mathrm{A}_x\mathrm{B}_{1-x}\mathrm{C}$), the parameters in the ABCM are calculated in the virtual crystal approximation (VCA).\cite{S-Abeles1963,S-Adachi1985}

Note that the full phonon dispersions are important in order to obtain an accurate density of phonon states, which results in an accurate thermal conductivity tensor. In Sec. S-II, we present a detailed analysis of the role of the full dispersion by comparing the calculations with best isotropic fits to those with full dispersion. The difference between the corresponding thermal conductivities $\kappa$ along a given direction can be appreciable (subscripts iso and full denote the type of dispersion): $\kappa_{\mathrm{iso}}\approx 0.6 \kappa_{\mathrm{full}}$ for binary compounds and $\kappa_{\mathrm{iso}}\approx 1.2 \kappa_{\mathrm{full}}$ for evenly mixed ternaries.

We only include acoustic phonon branches, which are the dominant carriers of heat in semiconductors and whose dispersions are accurately captured by the VCA.

\subsection{Phonon Interactions in Bulk III-V Compound Semiconductors} \label{sub:scatter}
The total phonon relaxation time of a phonon in branch b and with wavevector $\vec{q}$ is given by
\begin{equation}
\tau_\mathrm{b}^{-1}(\vec{q},T)=\sum_{i}\tau_{\mathrm{b},i}^{-1}(\vec{q},T),
\end{equation}
where $i$ represents the $i$th scattering mechanism. In this model, we consider all the important scattering mechanisms for thermal transport in III-Vs: the three-phonon umklapp (U) and normal (N) scattering processes (often referred to together simply as phonon-phonon scattering), mass-difference scattering (due to naturally occurring isotopes or alloying), and scattering with charge carriers and ionized dopants. The following subsections discuss each mechanisms, while Table \ref{tab:parameters} lists all the parameters used in the calculations.

\subsubsection{Phonon-phonon Scattering}
The three-particle phonon-phonon scattering is the dominant mechanism over a wide temperature range. We consider both U and N processes in our calculations.\cite{S-Ziman1960,S-Lindsay2013} In U processes, phonon momentum is not explicitly conserved, which leads to a thermal resistance. N processes do not cause thermal resistance directly, but rather influence the strength of U processes as they redistribute the phonons among branches. The relaxation time due to the U processes is\cite{S-Slack1964}
\begin{equation}
\tau_\mathrm{b,U}^{-1}(\vec{q},T)=\frac{\hbar\gamma_\mathrm{b}^2\omega_\mathrm{b}^2(\vec{q})}{\bar{M}v_b^2(\vec{q})\Theta_\mathrm{b,D}}Te^{-\Theta_\mathrm{b,D}/3T},
\end{equation}
where $\bar{M}$ is the average mass of an atom in the crystal, and $\gamma_\mathrm{b}$ is the Gr\"{u}neissen parameter for branch b. Since experimentally measured values of $\gamma_\mathrm{b}$ vary a lot,\cite{S-Soma1985,S-Soma1987,S-Talwar1990} we started with a reference value and slightly adjusted it to get the best fit to the experimentally obtained thermal conductivity (see Table \ref{tab:parameters}). $\Theta_\mathrm{b,D}$ is the mode-specific Debye temperature calculated from
\begin{equation}\label{equ:debye}
\Theta_\mathrm{b,D}^2=\frac{5\hbar^2}{3k_\mathrm{B}^2}\frac{\int\omega^2D_\mathrm{b}(\omega)~d\omega}{\int D_\mathrm{b}(\omega)~d\omega},
\end{equation}
where $D_\mathrm{b}(\omega)$ is the mode-specific phonon density of states (DOS) calculated from the full phonon dispersion using the numerical method proposed by Gilat and Raubenheimer.\cite{S-Gilat1966}
The relaxation time due to normal scattering has different forms for the transverse acoustic (TA) and the longitudinal acoustic (LA) phonons, and the form is given by \cite{S-Asen-Palmer1997,S-Morelli2002}
\begin{equation}\label{equ:normalRate}
\tau_\mathrm{b,N}^{-1}(\vec{q},T)=\begin{cases}
B_\mathrm{N}^T\omega_\mathrm{b}(\vec{q})T^4,~~~\mathrm{b}=TA, \\
B_\mathrm{N}^L\omega_\mathrm{b}^2(\vec{q})T^3,~~~\mathrm{b}=LA,
\end{cases}
\end{equation}
and
\begin{subequations}
	\begin{equation}
	B_\mathrm{N}^\mathrm{T} \sim\frac{k_\mathrm{B}^4\gamma_\mathrm{T}^2\Omega_0}{\bar{M}\hbar^3v_\mathrm{T}^5(\vec{q})},\\
	\end{equation}
	\begin{equation}
	 B_\mathrm{N}^\mathrm{L}
	 \sim\frac{k_\mathrm{B}^3\gamma_\mathrm{L}^2\Omega_0}{\bar{M}\hbar^2v_\mathrm{L}^5(\vec{q})}.
	\end{equation}
\end{subequations}
Here, $\Omega_0$ is the average volume occupied by an atom in the lattice. We assume that the same Gr\"{u}neissen parameter holds for both N and U processes in a given branch. Given that the high-temperature limit to thermal conductivity is dominated by N and U processes, we tune the branch Gr\"{u}neissen parameter until the best fit to the high-temperature ($>$ 300 K) thermal conductivity is found. In our calculations, $B_\mathrm{N}^\mathrm{T}$ is on the order of $10^{-14}~\mathrm{s}^{-1}\mathrm{K}^{-3}$ and $B_\mathrm{N}^\mathrm{L}$ is on the order of $10^{-26}~\mathrm{s}^{-1}\mathrm{K}^{-5}$, varying slightly as the group velocity and mass difference changes for different materials (see Table \ref{tab:parameters}).
\subsubsection{Mass-difference Scattering}
There are two major sources of mass-difference scattering in compound semiconductors: the natural occurrence of isotopes and the fact that the compound is formed with different elements. The scattering rate of a phonon due to mass-difference is\cite{S-Tamura1983,S-Maris1990}
\begin{equation}
\tau_\mathrm{b,M}^{-1}(\vec{q})=\frac{\pi}{6}\Omega_0\Gamma_\mathrm{M}\omega_\mathrm{b}^2(\vec{q})D_\mathrm{b}(\omega_\mathrm{b}(\vec{q})),
\end{equation}
where $\Gamma_\mathrm{M}$ is the total mass parameter, obtained by summing all the mass parameters.

For a single element $A$ with an average mass of $\bar{M_\mathrm{A}} =\sum\limits_{i}f_\mathrm{i}M_\mathrm{i,A}$, where $f_\mathrm{i}$ and $M_\mathrm{i,A}$ are the abundance and the mass of the $i$th isotope of element A, the isotope mass parameter is
\begin{equation}
\Gamma_\mathrm{I}(\mathrm{A})=\sum\limits_{i}f_\mathrm{i}(1-M_\mathrm{i,A}/\bar{M_\mathrm{A}})^2.
\end{equation}
For a compound with the form of $\mathrm{A}_x\mathrm{B}_y\mathrm{C}_z...$, the effective isotope mass parameter can be expressed as\cite{S-Holland1964}
\begin{subequations}
	\begin{equation}
	\Gamma_\mathrm{iso}=\frac{x\left(\frac{\bar{M_\mathrm{A}}}{\bar{M}}\right)^2\Gamma_\mathrm{I}(\mathrm{A})+y\left(\frac{\bar{M_\mathrm{B}}}{\bar{M}}\right)^2\Gamma_\mathrm{I}(\mathrm{B})
+z\left(\frac{\bar{M_\mathrm{C}}}{\bar{M}}\right)^2\Gamma_\mathrm{I}(\mathrm{C})\ldots}{x+y+z+\ldots},
	\end{equation}
where
	\begin{equation}
	\bar{M}=\frac{x\bar{M_\mathrm{A}}+y\bar{M_\mathrm{B}}+z\bar{M_\mathrm{C}}+\ldots}{x+y+z+\ldots}
	\end{equation}
is the average atom mass in the compound.

Here, we are interested in both binary and ternary III-As compounds, so the general form of the compound becomes $\mathrm{A}_x\mathrm{B}_{1-x}\mathrm{As},~(0\leq x\leq 1)$. Since the element As has only one stable isotope, $^{75}\mathrm{As}$ (Ref.~\onlinecite{Shore2010}), our total isotope mass parameter simplifies to
	\begin{equation}
	\Gamma_\mathrm{iso}=2\frac{x\bar{M_\mathrm{A}}^2\Gamma_\mathrm{I}(A)+(1-x)\bar{M_\mathrm{B}}^2\Gamma_\mathrm{I}(B)}{\left[x\bar{M_\mathrm{A}}+(1-x)\bar{M_\mathrm{B}}+M_\mathrm{As}\right]^2}.
	\end{equation}
\end{subequations}

In binary $(x=0$ or $x=1)$ compounds, isotope scattering is dominant at low temperatures ($<150$ K). In ternary $(0<x<1)$ compounds, different masses of the group-III elements $A$ and $B$ result in alloy mass-difference scattering. The alloy mass-difference parameter is\cite{S-Abeles1963,S-Adachi1983}
\begin{subequations}
	\begin{equation}
	\Gamma_\mathrm{alloy}=x(1-x)\left[(\Delta M/\bar{M_\mathrm{III}})^2+\epsilon(\Delta a/\bar{a})^2\right],
	\end{equation}
where
	\begin{align}
	&\Delta M = \bar{M_\mathrm{A}}-\bar{M_\mathrm{B}},\\
	&\bar{M_\mathrm{III}} =x\bar{M_\mathrm{A}}+(1-x)\bar{M_\mathrm{B}},\\
	&\Delta a = a_\mathrm{A}-a_\mathrm{B},\\
	&\bar{a} =xa_\mathrm{A}+(1-x)a_\mathrm{B}.
	\end{align}
\end{subequations}
Here, $a_\mathrm{A}$ and $a_\mathrm{B}$ are the lattice constant of binary compounds AAs and BAs, respectively. $\epsilon$ is an empirically determined constant, which captures the scattering caused by the mismatch of lattice constants. We take $\epsilon=45$ for the III-V compounds following Abeles.\cite{S-Abeles1963}

Relative abundance of isotopes is taken as: $^{26}$Al: 0.001, $^{27}$Al: 0.999; $^{71}$Ga: 0.3989, $^{69}$Ga: 0.6011; $^{113}$In: 0.0429, $^{115}$In: 0.9571.
Assumed \textit{n}-type dopant is Si, $\bar{M}_{\mathrm{Si}}$=28.085. Other parameters are in Table \ref{tab:parameters}.

\begin{table}
	\begin{tabular}{ l@{\hspace{0.5em}}c@{\hspace{0.7em}}c@{\hspace{0.7em}}c }
		\hline\hline
		 Material & GaAs & InAs & AlAs \\
		\hline
		$\gamma_{\mathrm{TA1}}$ (exp.) & 0.57 & 0.58 & 0.46 \\
        $\gamma_{\mathrm{TA1}}$ (\textit{ab initio}) & 0.52 & 0.46 & 0.46 \\

        $\gamma_{\mathrm{TA2}}$ (exp.) & 0.57 & 0.58 & 0.46 \\
        $\gamma_{\mathrm{TA2}}$ (\textit{ab initio}) & 0.52 & 0.46 & 0.46 \\

        $\gamma_{\mathrm{LA}}$ (exp.) & 1.35 & 1.6 & 1.35 \\
        $\gamma_{\mathrm{LA}}$ (\textit{ab initio}) & 1.3 & 1.35 & 1.35 \\

        \hline

        $\Theta_{\mathrm{TA1,D}}$ (K) & 141.21 & 103.93 & 181.59 \\
        $\Theta_{\mathrm{TA2,D}}$ (K) & 167.76 & 124.64 & 215.21 \\		
        $\Theta_{\mathrm{LA,D}}$ (K) & 304.96 & 253.16 & 380.07 \\

        \hline

		$B^{\mathrm{TA1}}_{\mathrm{N}}$ ($10^{-14}\,\mathrm{s}^{-1}\mathrm{K}^{-3}$) & 0.421 & 1.33 & 0.166 \\
        $B^{\mathrm{TA2}}_{\mathrm{N}}$ ($10^{-14}\,\mathrm{s}^{-1}\mathrm{K}^{-3}$) & 0.443 & 1.39 & 0.173 \\
        $B^{\mathrm{LA}}_{\mathrm{N}}$ ($10^{-26}\,\mathrm{s}^{-1}\mathrm{K}^{-5}$) & 1.27 & 2.85 & 7.04 \\

        \hline

		$v_{\mathrm{TA1}}$ (m/s)& 3397.3 & 2817.5 & 4318.2 \\
        $v_{\mathrm{TA2}}$ (m/s)& 3361.9 & 2793.0 & 4284.7 \\
        $v_{\mathrm{LA}}$ (m/s)& 5418.2 & 4708.7 & 6812.4 \\

        \hline
        $a_0$ (\AA)& 5.6532 & 6.0583 & 5.6611 \\
        $\Omega_0=a_0^3/4$ ($10^{-29}$ m)& 4.5168 & 5.5589 & 4.5357 \\
        \hline
        $m^{*}$ ($m_0=9.1\times 10^{-31}$ kg) & 0.063 & 0.023 & 0.1\\
        $\bar{M}$ (a.u.=$1.66054\times 10^{-27}$ kg) & 69.7978 & 144.9142  & 26.999\\
        $\rho$ ($10^3$ kg/m$^3$) &5.3232 &5.673 & 3.7342\\
		\hline\hline
	\end{tabular}
	\caption{Materials parameters used in the calculation.}\label{tab:parameters}
\end{table}

\subsubsection{Dopant and Electron Scattering}

When III-V compound semiconductors are doped (we consider \textit{n}-type only for our applications of interest), group-III atoms may be randomly replaced by a group IV dopant, creating extra free electrons. Doping introduces two additional scattering mechanisms, phonon-electron interaction and impurity mass-difference scattering. At low doping levels, the relaxation time due to phonon-electron interaction can be expressed as\cite{S-Parrott1979}
\begin{equation}
\tau_\mathrm{b,ph-e}^{-1}(\vec{q})=\frac{N_\mathrm{D}\xi_\mathrm{def}^2}{\rho v_\mathrm{b}^2(\vec{q})k_\mathrm{B}T}\sqrt{\frac{\pi m^*v_\mathrm{b}^2(\vec{q})}{2k_\mathrm{B}T}}exp\left(-\frac{m^*v_\mathrm{b}^2(\vec{q})}{2k_\mathrm{B}T}\right).
\end{equation}
Here, $N_\mathrm{D}$ represents the doping concentration, $\xi_\mathrm{def}$ is the deformation potential, $\rho$ is the density of the crystal, and $m^*$ is the electron effective mass in the crystal. For ternary materials, $m^*$ is obtained from a weighted average of those in the constituent binary materials.\cite{S-Adachi1985}

The impurity mass-difference scattering yields an additional contribution to the total mass-difference parameter discussed in the previous section,\cite{S-Ramsbey1992}
\begin{equation}
\Gamma_\mathrm{imp}=\frac{N_\mathrm{D}}{N_0}\left(\frac{\delta M}{\bar{M}}\right)^2.
\end{equation}
$N_0=1/\Omega_0$ is the concentration of native atoms, and $\delta M=M_\mathrm{D}-\bar{M}_\mathrm{III}$ is the mass difference between the dopant and the average mass of group III atoms being replaced.

The effect of scattering introduced by doping is negligible when the impurity density is below $10^{17}~\mathrm{cm}^{-3}$, and still small compared to other scattering rates when the doping density exceeds $10^{18}~\mathrm{cm}^{-3}$ , in good agreement with experiment.\cite{S-Amith1965}

\subsection{Benchmark: Thermal Conductivity of Binary Compounds} \label{sub:binary}

We apply the scattering rates and the thermal conductivity model to three binary arsenide compounds -- GaAs, AlAs, and InAs -- whose thermal properties over a wide temperature range have been extensively studied both experimentally\cite{S-Amith1965,S-Carlson1965,S-Inyushkin2003,S-Afromowitz1973,S-Adachi1993,S-Evans2008,S-Bowers1959,S-Tamarin1971,S-Guillou1972,S-Heckman2008} and via \textit{ab inito} calculations.\cite{S-Lindsay2013}

\begin{figure}
    \includegraphics[width = 3 in]{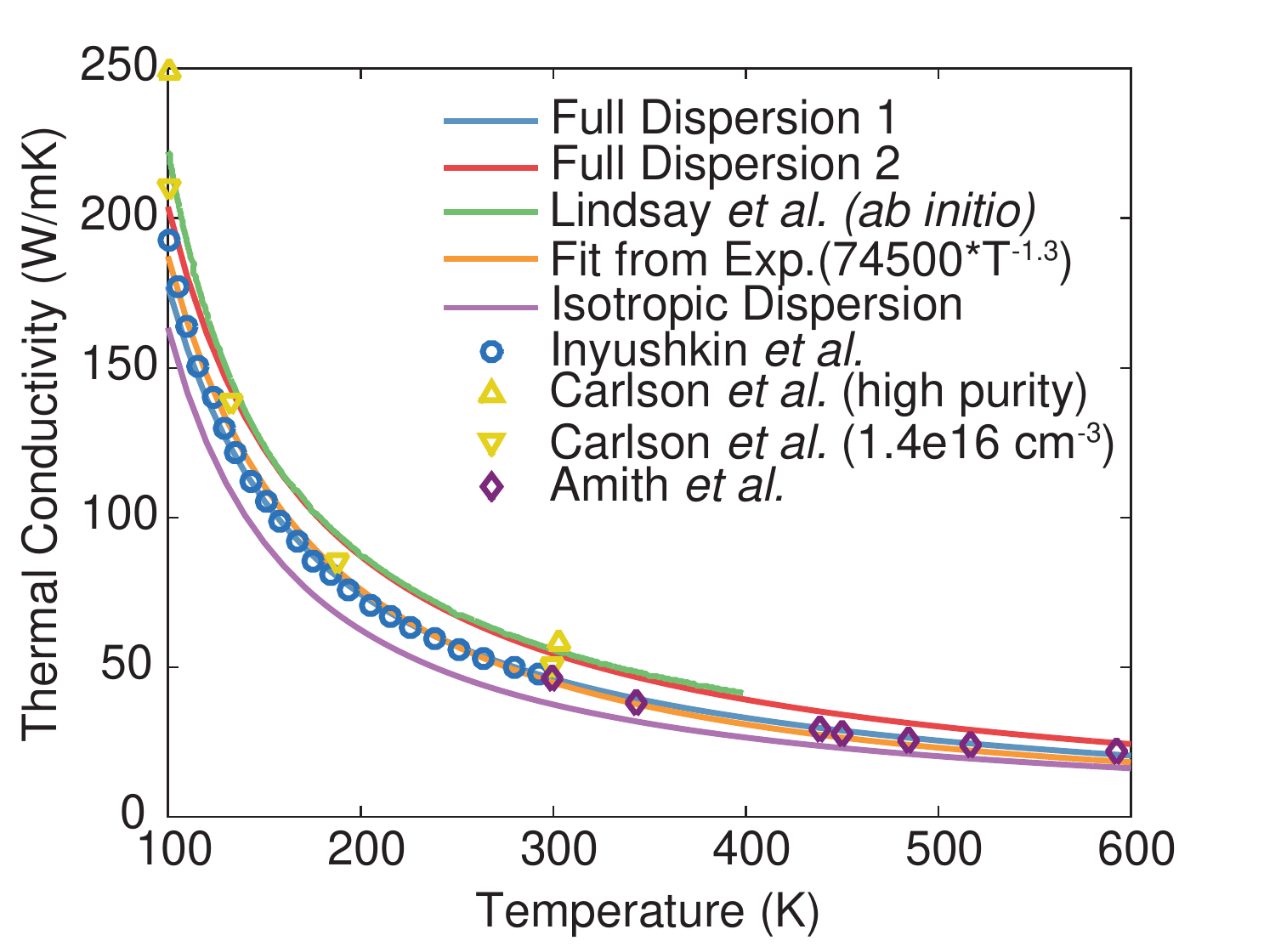}
	\caption{Thermal conductivity of bulk GaAs based on our calculation with full dispersion; $\gamma_\mathrm{TA}=0.57,~\gamma_\mathrm{LA}=1.35$ (blue curve) and $\gamma_\mathrm{TA}=0.52,~\gamma_\mathrm{LA}=1.30$ (red curve). The green curve shows the \textit{ab initio} results from Lindsay \textit{et al.}.\cite{S-Lindsay2013} The orange curve is an analytic fit to the experimental data of Evans \textit{et al}.\cite{S-Evans2008} The purple curve is the calculated thermal conductivity based on our model and with the isotropic dispersion approximation (see Appendix \ref{sec:isodisp}). Blue circles, yellow triangles, and blue diamonds correspond to the experimental data from Inyushkin \textit{et al.},\cite{S-Inyushkin2003} Carlson \textit{et al.},\cite{S-Carlson1965} and Amith \textit{et al.},\cite{S-Amith1965} respectively.}
	\label{fig:GaAsbulk}
\end{figure}

Figure~\ref{fig:GaAsbulk} shows the thermal conductivity of GaAs obtained from the full dispersion calculation from our model in comparison with the experimental results from Amith~\textit{et al.},\cite{S-Amith1965} Carlson~\textit{et al.},\cite{S-Carlson1965} Inyushkin~\textit{et al.},\cite{S-Inyushkin2003} and the \textit{ab inito} results of Lindsay~\textit{et al.}\cite{S-Lindsay2013} By slightly adjusting the Gr\"{u}neissen parameter, we can get thermal conductivity that agrees very well with either experiment (green curve) or first-principles calculations (light-blue curve).

\begin{figure}
    \includegraphics[width = 3 in]{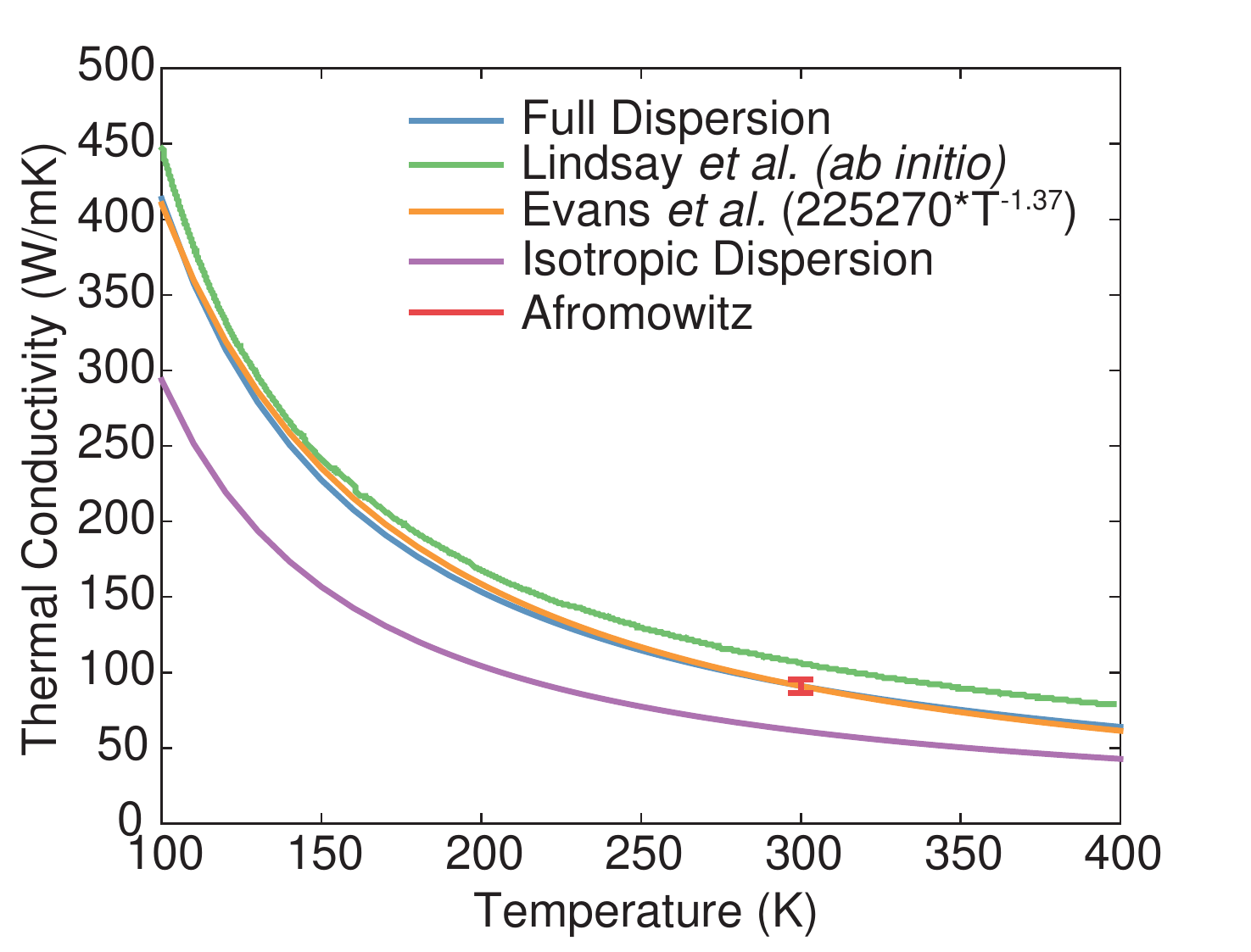}
	\caption{Thermal conductivity of bulk AlAs based on our calculation with full dispersion. $\gamma_\mathrm{TA}=0.46$ and $\gamma_\mathrm{LA}=1.35$ (blue curve). In green, orange, and purple we show the \textit{ab initio} data from Lindsay \textit{et al.},\cite{S-Lindsay2013} an analytic fit to the experimental data from Evans \textit{et al.},\cite{S-Evans2008} and the calculation with isotropic dispersion approximation (see Appendix \ref{sec:isodisp}), respectively. The red symbol shows experimental data from Afromowitz.\cite{S-Afromowitz1973}}
	\label{fig:AlAsbulk}
\end{figure}

\begin{figure}
\includegraphics[width = 3 in]{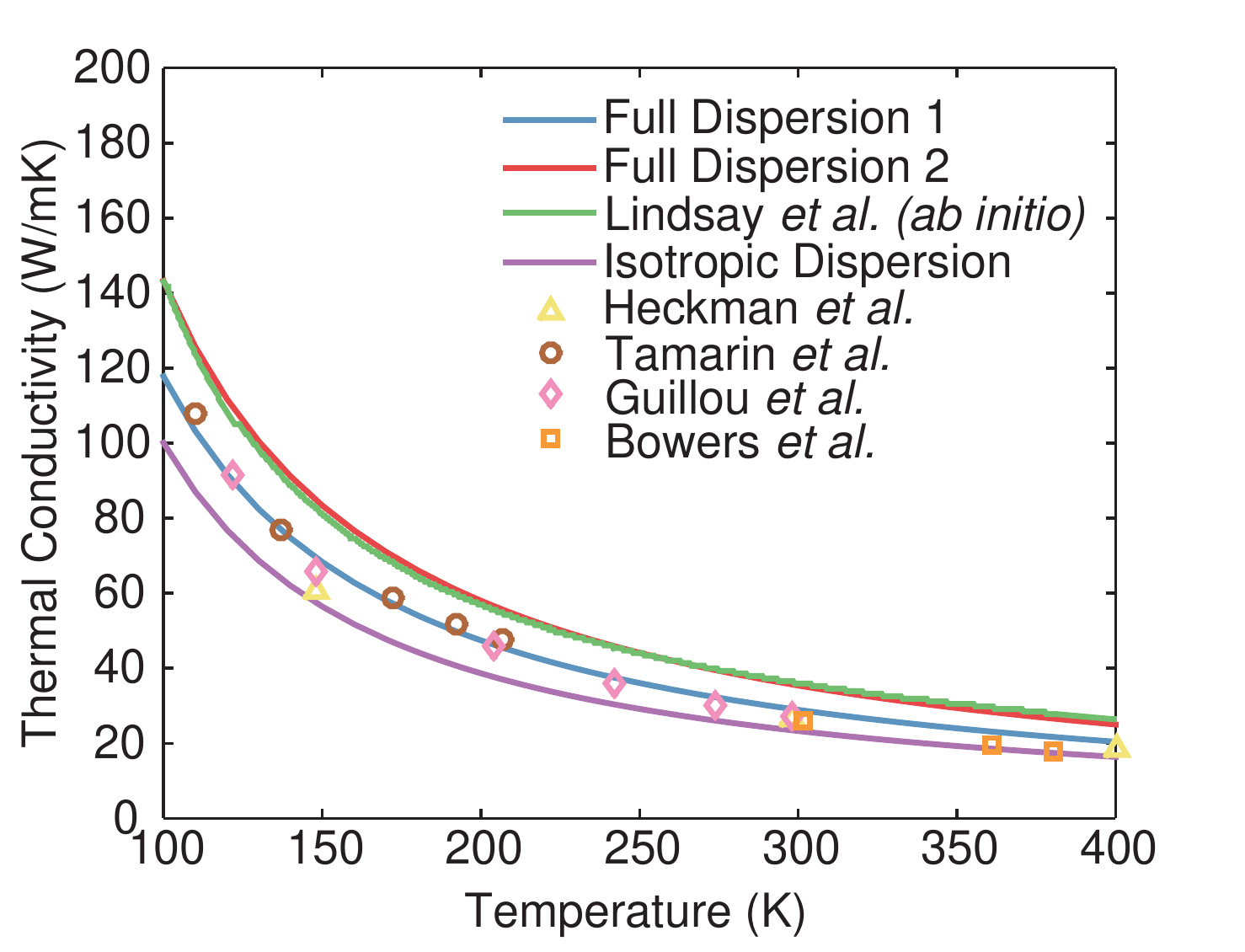}
	\caption{Thermal conductivity of bulk InAs based on our calculation with full dispersion; $\gamma_\mathrm{TA}=0.56,~\gamma_\mathrm{LA}=1.60$ (blue curve) and $\gamma_\mathrm{TA}=0.46,~\gamma_\mathrm{LA}=1.35$ (red curve). The green curve shows the \textit{ab initio} results from Lindsay \textit{et al.}\cite{S-Lindsay2013} The purple curve shows the calculated thermal conductivity with the isotropic dispersion approximation. Yellow triangles, brown circles, pink diamonds, and orange squares correspond to  the experimental data from Heckman \textit{et al.},\cite{S-Heckman2008} Tamarin \textit{et al.},\cite{S-Tamarin1971} Guillou \textit{et al.},\cite{S-Guillou1972} and Bowers \textit{et al.},\cite{S-Bowers1959} respectively.}
	\label{fig:InAsbulk}
\end{figure}

Figures~\ref{fig:AlAsbulk} and \ref{fig:InAsbulk} show similar comparisons for the thermal conductivity of AlAs\cite{S-Afromowitz1973,S-Adachi1993,S-Evans2008,S-Lindsay2013} and InAs,\cite{S-Bowers1959,S-Tamarin1971,S-Guillou1972,S-Heckman2008,S-Lindsay2013} respectively.


\subsection{Benchmark: Thermal Conductivity of Ternary Compounds} \label{sub:ternary}

\begin{figure}
\includegraphics[width = 3 in]{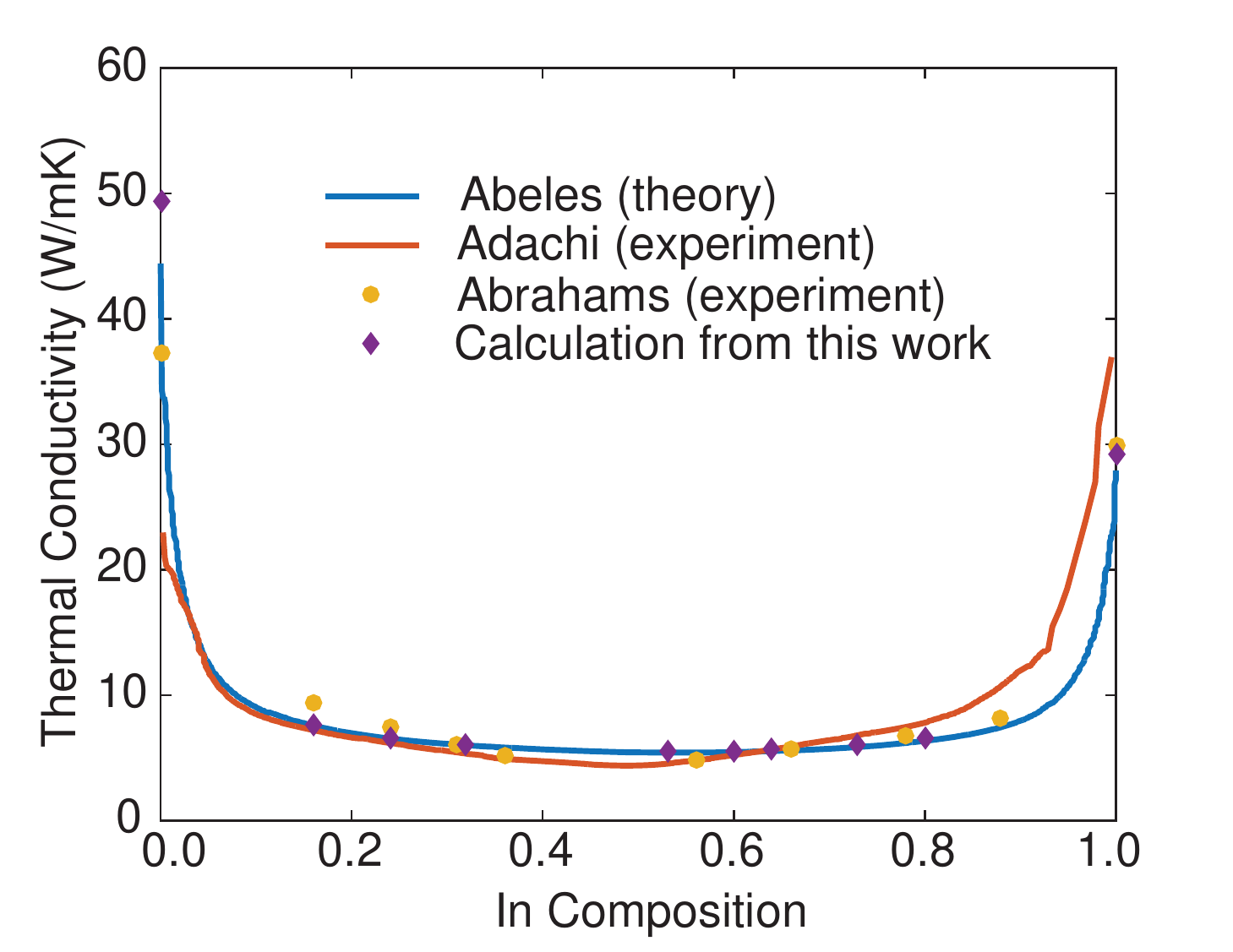}
	\caption{Thermal conductivity of bulk $\mathrm{In}_x\mathrm{Ga}_{1-x}\mathrm{As}$ with varying In composition. The blue curve shows the theoretical results from Abeles.\cite{S-Abeles1963} The red curve and orange dots present the experimental results from Adachi\cite{S-Adachi1983} and Abrahams \textit{et al.},\cite{S-Abrahams1959} respectively. Purple diamonds represent the results of our calculation.}
	\label{fig:InGaAs}
\end{figure}

\begin{figure}
\includegraphics[width = 3 in]{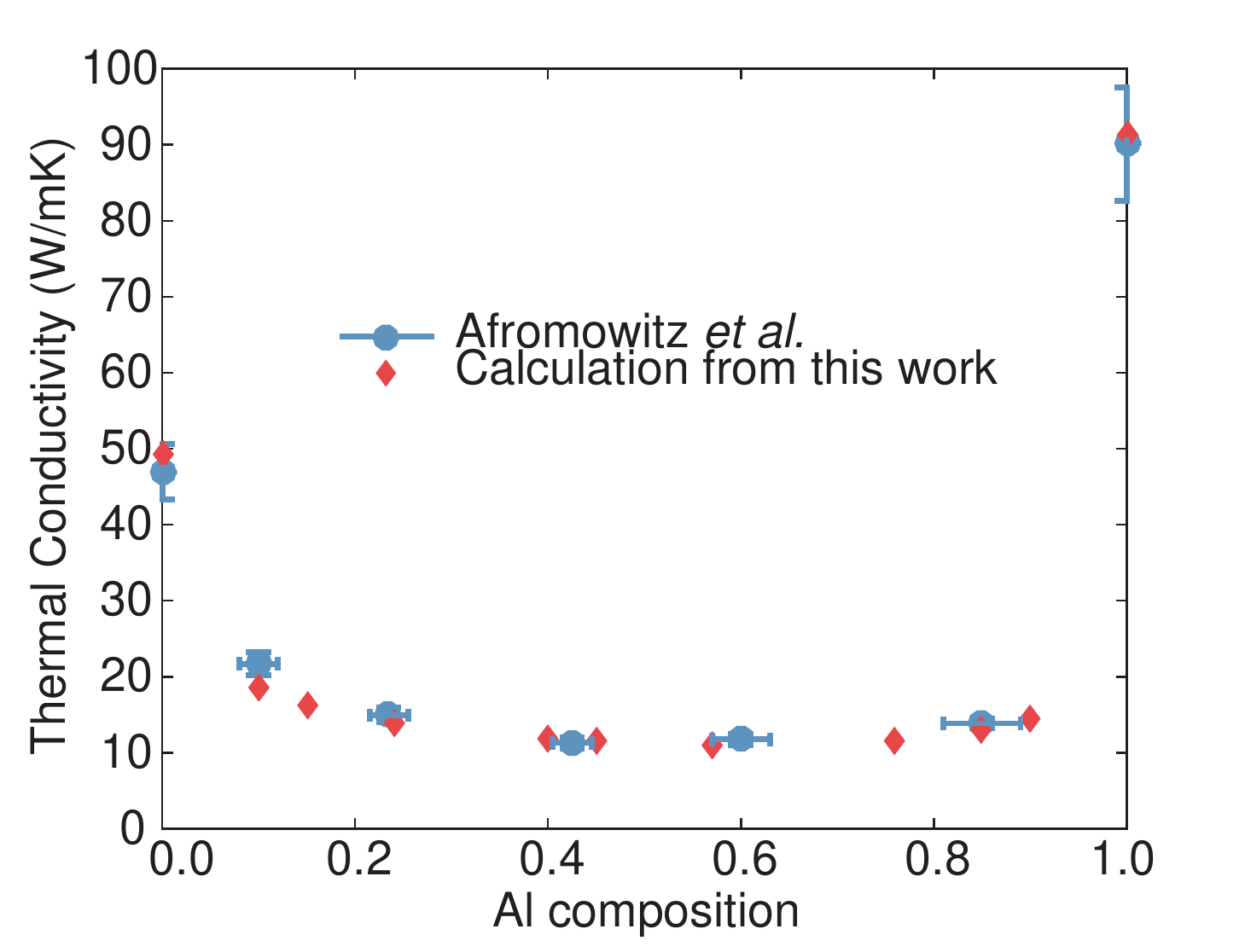}
	\caption{Thermal conductivity of bulk $\mathrm{Al}_x\mathrm{Ga}_{1-x}\mathrm{As}$ with varying Al composition. Blue dots and red diamonds show the experimental data of  Afromowitz \textit{et al.}\cite{S-Afromowitz1973} and the results of our calculation, respectively.}
	\label{fig:AlGaAs}
\end{figure}

Alloy scattering is the dominant mechanism that influences the thermal conductivity of ternary arsenide compounds. Unfortunately, very few experiments\cite{S-Abrahams1959,S-Abeles1963,S-Afromowitz1973,S-Adachi1983} have been carried out on these materials and they were all performed at room temperature. In Figures~\ref{fig:InGaAs} and \ref{fig:AlGaAs}, we compare our calculated thermal conductivity of $\mathrm{In}_x\mathrm{Ga}_{1-x}\mathrm{As}$ and $\mathrm{Al}_x\mathrm{Ga}_{1-x}\mathrm{As}$ with various compositions with experimental results.\cite{S-Abrahams1959,S-Abeles1963,S-Afromowitz1973,S-Adachi1983} Our calculations agree well with the available experimental data over a wide range of compositions. To our knowledge, no systematic measurements of the thermal conductivity of $\mathrm{In}_x\mathrm{Al}_{1-x}\mathrm{As}$ have been carried out thus far.


\section{Parameterized Isotropic Dispersion} \label{sec:isodisp}

\begin{figure}
\includegraphics[width = 3 in]{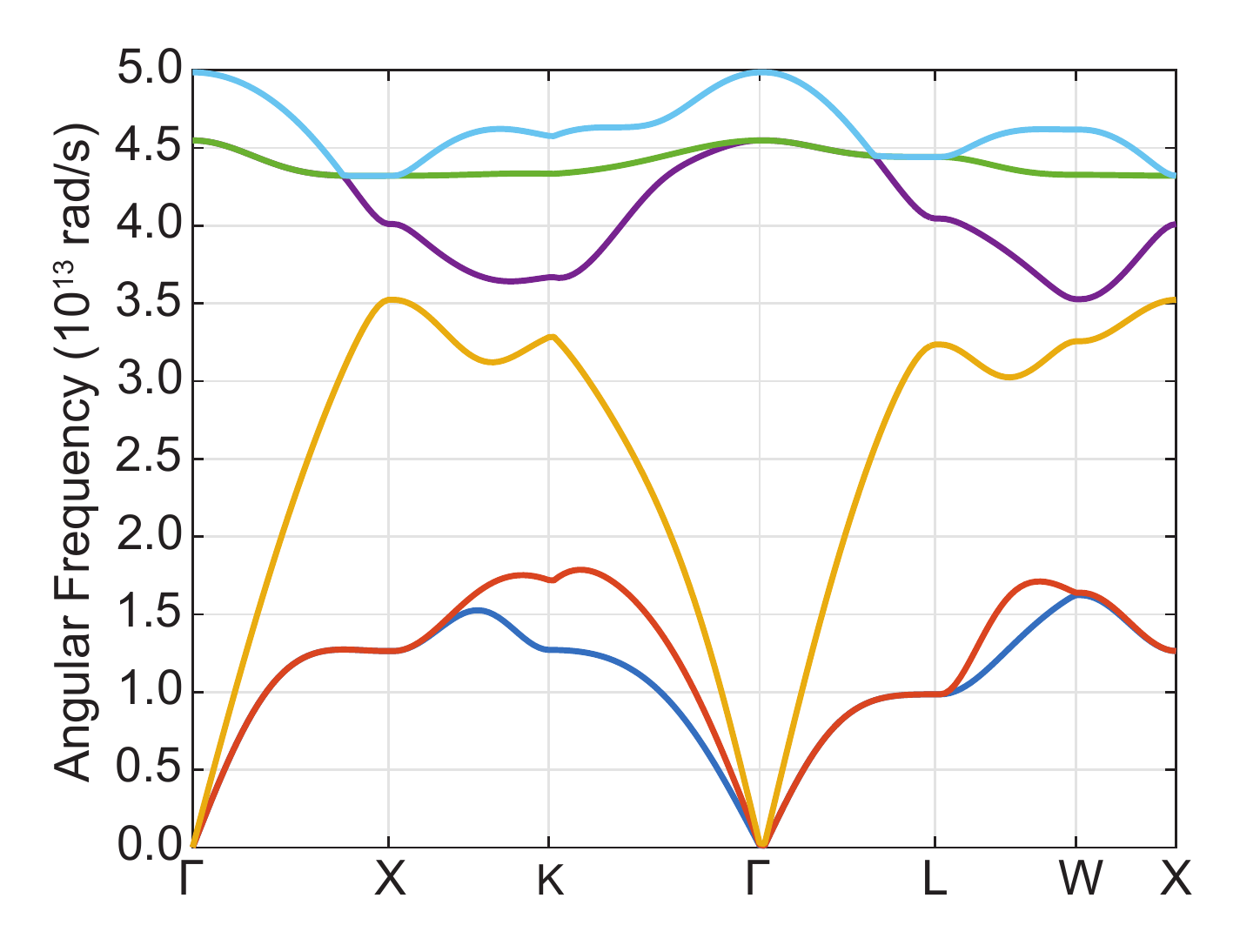}
	\caption{Typical phonon dispersions of a ternary group-III arsenide compound ($\mathrm{In}_{0.53}\mathrm{Ga}_{0.47}\mathrm{As}$) along high-symmetry directions in the Brillouin zone.}
	\label{fig:fulldisp}
\end{figure}

In order to calculate the thermal conductivity with full dispersion relations, we need to calculate and store the information for each material, which requires a lot of computation time and memory. However, it is a necessity in our case to achieve accuracy. First, we want to be able to calculate the thermal conductivity of a ternary material with any given composition for which experimental work may generally not be available. Besides, experiments cannot give us the dispersion information on any composition and wave vector we want. Figure~\ref{fig:fulldisp} shows the exact phonon dispersion of $\mathrm{In}_{0.53}\mathrm{Ga}_{0.47}\mathrm{As}$ along highly symmetric directions calculated from the ABCM. As we can see, the dispersion is not isotropic, and two TA branches are degenerate along the [100] ($\Gamma-X$) direction. Also, we can see that the isotropic Debye approximation or the sine approximation of the dispersion are not ideal in capturing the features.

\begin{figure}
\includegraphics[width = 3 in]{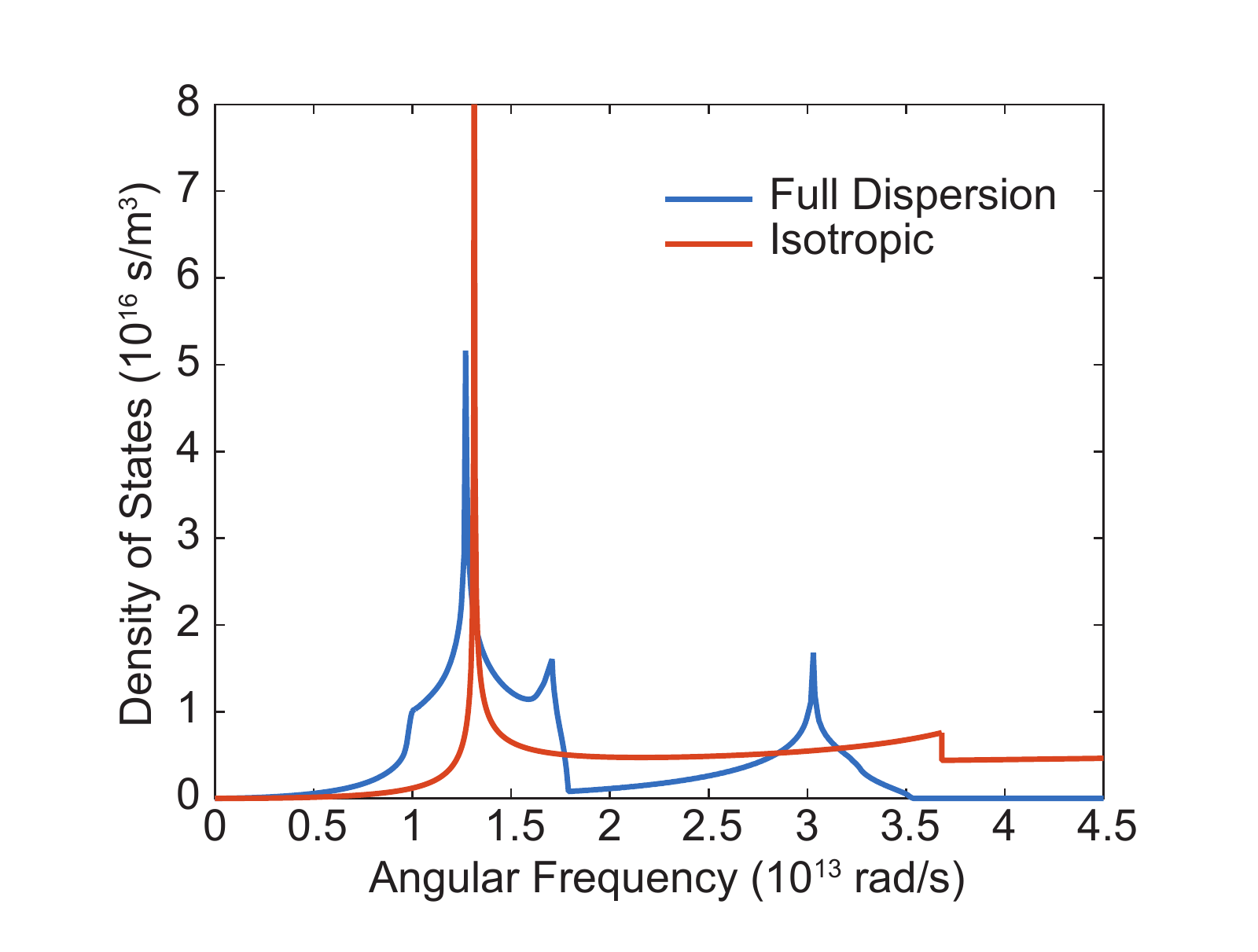}
	\caption{The phonon density of states for $\mathrm{In}_{0.53}\mathrm{Ga}_{0.47}\mathrm{As}$ calculated based on full dispersions and from an isotropic fit.}
	\label{fig:cmpDOS}
\end{figure}

Thermal conductivity calculation of the three binary compounds (Figs. \ref{fig:GaAsbulk}-\ref{fig:InAsbulk}) show that using isotropic dispersions underestimates thermal conductivity for the binaries, primarily because the two TA branches are not actually degenerate (Fig.~\ref{fig:fulldisp}), and TA2 carries more heat than TA1 because of the higher average group velocity. The isotropic dispersion also overestimates the thermal conductivity of ternary materials, mainly because the phonon DOS calculated from the isotropic dispersion is much smaller than that from full dispersion (Fig.~\ref{fig:cmpDOS}).

\begin{figure}
\includegraphics[width = 3 in]{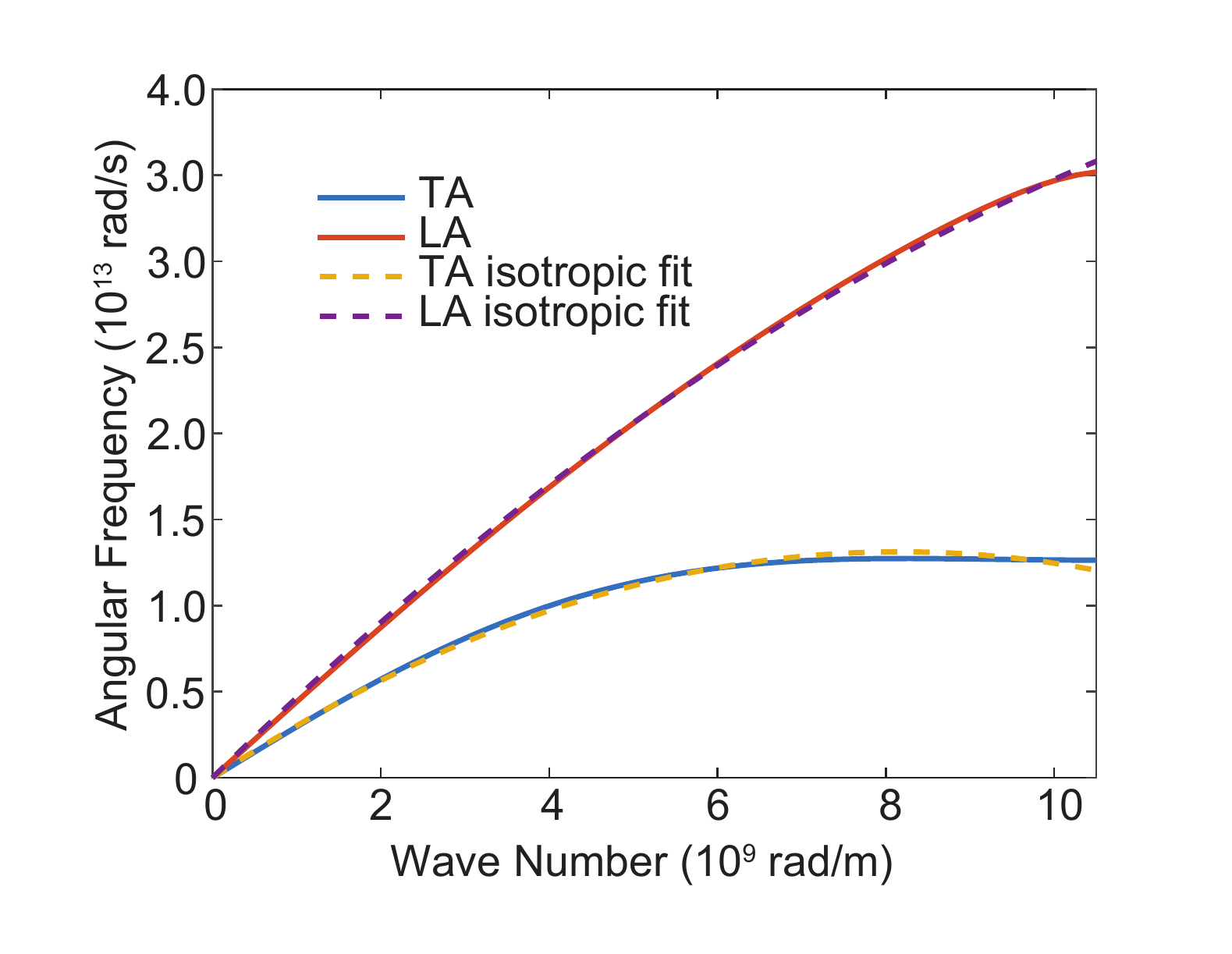}
	\caption{Comparison of the full dispersion with the isotropic fit (\ref{eq:isotropic}) along the [100] direction for $\mathrm{In}_{0.53}\mathrm{Ga}_{0.47}\mathrm{As}$.}
	\label{fig:fitdisp}
\end{figure}

\begin{figure}
\includegraphics[width = 3 in]{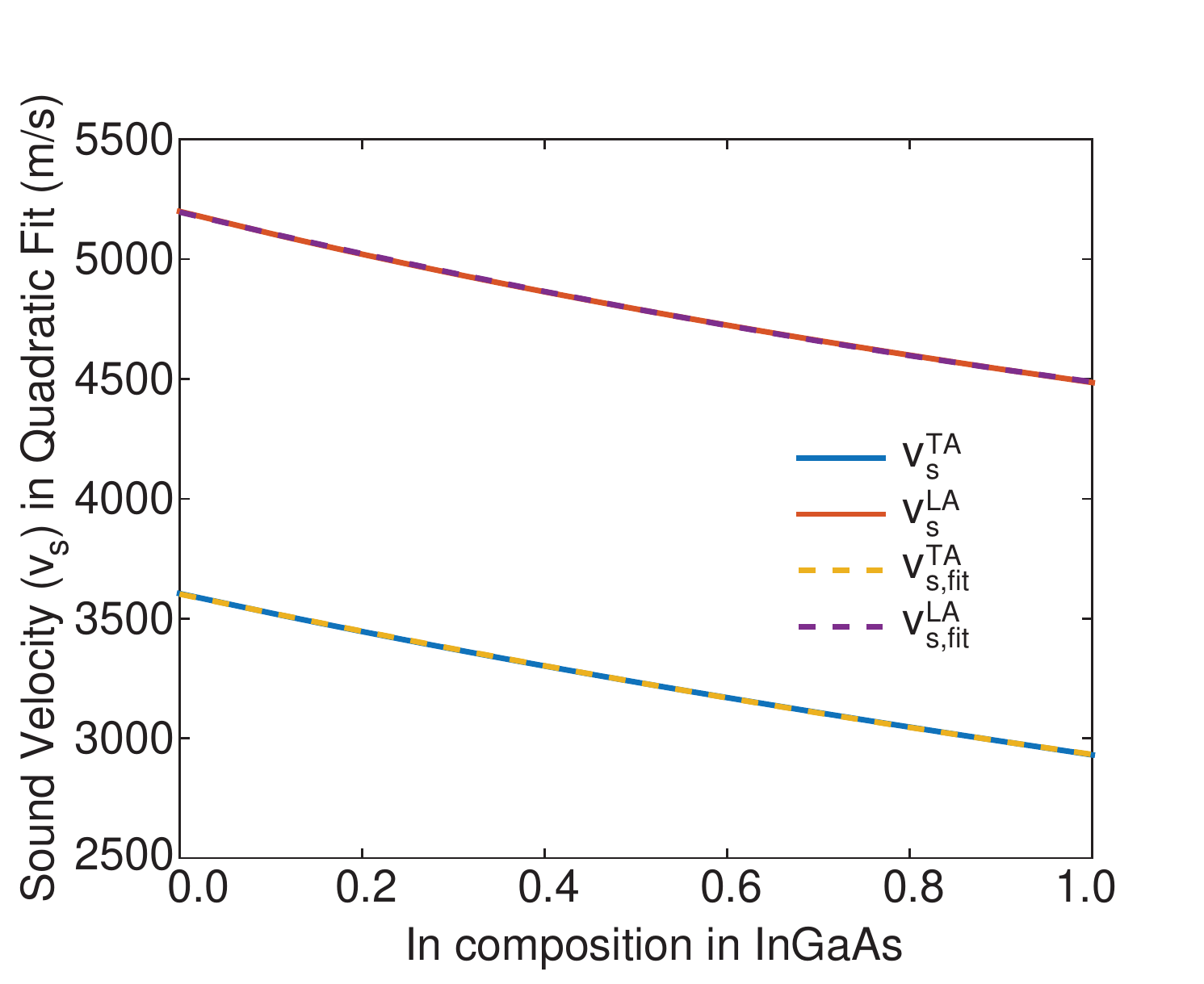}
	\caption{Sound velocity in the isotropic fit (\ref{eq:isotropic}) to the full dispersion of InGaAs along the [100] direction as a function of In content.}
	\label{fig:isovs}
\end{figure}

\begin{figure}
\includegraphics[width = 3 in]{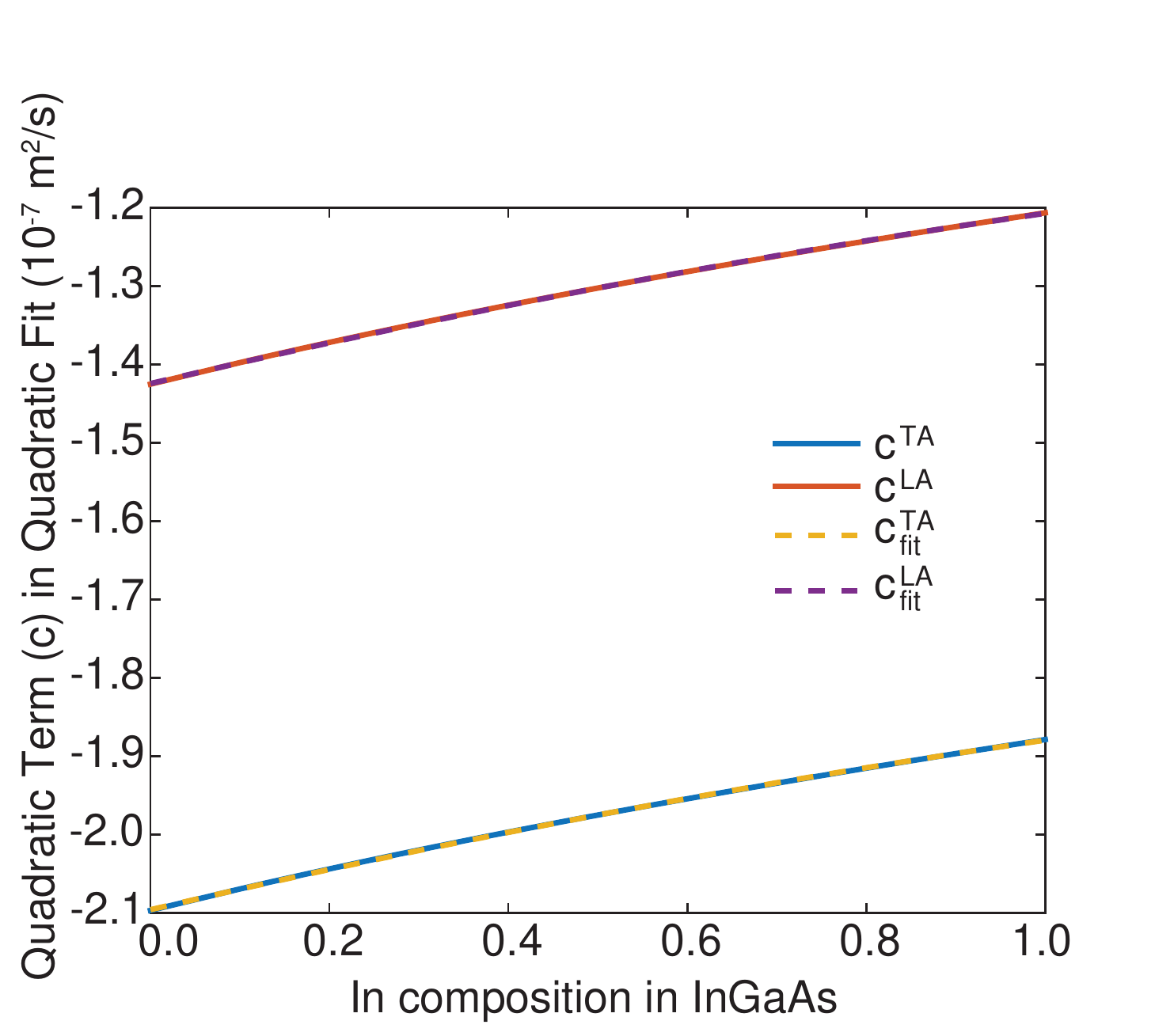}
	\caption{Quadratic term in the isotropic fit (\ref{eq:isotropic}) to the full dispersion dispersion of InGaAs along the [100] direction as a function of In content.}
	\label{fig:isoc}
\end{figure}

To take advantage of the work we have done, and make it easier to get a sense of what the dispersion relation of a ternary compounds with a random composition is like, we fit our full dispersion data along [100] direction with a quadratic expression, which has been shown to perform well in materials with similar crystal structures, such as Si\cite{S-Maurer2015} and GaN.\cite{S-Davoody2014}
With the two TA branches being degenerate, the dispersion relation reads
\begin{equation}\label{eq:isotropic}
\omega_\mathrm{b}(q)=\begin{cases}
v_\mathrm{s}^\mathrm{T}q+c^\mathrm{T}q^2,~~~\mathrm{b}=TA\\
v_\mathrm{s}^\mathrm{L}q+c^\mathrm{L}q^2,~~~\mathrm{b}=LA,
\end{cases}
\end{equation}
where $v_\mathrm{s}^\mathrm{T}$ and $v_\mathrm{s}^\mathrm{L}$ are the sound velocity of the two branches and $c^\mathrm{T}$ and $c^\mathrm{L}$ are the corresponding quadratic coefficients. Figure~\ref{fig:fitdisp} shows that the expressions yield a good fit to the dispersion.
We report the parameters in our quadratic fit, so that one can get an easy and reasonably accurate estimate of the phonon dispersion of ternary group-III arsenide materials with any composition. Figures~\ref{fig:isovs} and \ref{fig:isoc} show the two fitting parameters of InGaAs as a function of In composition. We find that both $v_\mathrm{s}$ and $c$ fit well to a quadratic expression
\begin{subequations}
	\begin{eqnarray}
	v_\mathrm{s}^\mathrm{b}(x) =& & v_\mathrm{s2}^\mathrm{b}x^2+v_\mathrm{s1}^\mathrm{b}x+v_\mathrm{s0}^\mathrm{b}; \\
	c^\mathrm{b}(x) =& & c_{2}^\mathrm{b}x^2+c_{1}^\mathrm{b}x+c_{0}^\mathrm{b},
	\end{eqnarray}
\end{subequations}
where $\mathrm{b}=$TA or LA.

We have calculated the parameters for three types of ternary compounds, $\mathrm{In}_x\mathrm{Ga}_{1-x}\mathrm{As}$, $\mathrm{In}_x\mathrm{Al}_{1-x}\mathrm{As}$, and $\mathrm{Al}_x\mathrm{Ga}_{1-x}\mathrm{As}$, and they are reported in Table~\ref{tab:isodisp}. The isotropic approximation (\ref{eq:isotropic}) to the full dispersion gives a fairly good estimate of the sound velocity. The calculated thermal conductivity based on the isotropic approximation does differ from that calculated with full dispersion: $\kappa_\mathrm{iso}\approx0.6\kappa_\mathrm{full}$ for binaries and $\kappa_\mathrm{iso}\approx1.2\kappa_\mathrm{full}$ for almost evenly mixed ternaries. The error is in between when a ternary is not an even mixed of two binaries. However, if high accuracy is not critical, good estimates are possible with isotropic dispersions that use the parameters given in Table \ref{tab:isodisp}.
\begin{table}
	\begin{tabular}{ c@{\hspace{0.5em}}c@{\hspace{0.7em}}c@{\hspace{0.7em}}c }
		\hline\hline
		 Material & InGaAs & InAlAs & AlGaAs \\
		\hline
		$v_\mathrm{s2}^\mathrm{T}$ & 130.50 & 116.59 & 555.58 \\
		$v_\mathrm{s1}^\mathrm{T}$ & -800.60 & -275.19 &422.04 \\
		$v_\mathrm{s0}^\mathrm{T}$ & 3602.3 & 4553.7 & 3616.6 \\
		$c_2^\mathrm{T}$ & -5$\times 10^{-9}$ & -6.2$\times 10^{-8}$ & -3.2$\times 10^{-8}$ \\
		$c_1^\mathrm{T}$ & 2.7$\times 10^{-8}$ & 1.3$\times 10^{-7}$ & -2.4$\times 10^{-8}$ \\
		$c_0^\mathrm{T}$ & -2.1$\times 10^{-7}$ & -2.6$\times 10^{-7}$ & -2.1$\times 10^{-7}$ \\
		\hline
		$v_\mathrm{s2}^\mathrm{L}$ & 199.94 & 165.43 & 782.56 \\
		$v_\mathrm{s1}^\mathrm{L}$ & -908.6 & -359.84 & 537.55 \\
		$v_\mathrm{s0}^\mathrm{L}$ & 5196.3 & 6480.0 & 5216.3 \\
		$c_2^\mathrm{L}$ & -5.3$\times 10^{-9}$ & -4.7$\times 10^{-8}$ & -2.3$\times 10^{-8}$ \\
		$c_1^\mathrm{L}$ & 2.7$\times 10^{-8}$ & 1.1$\times 10^{-7}$ & -1.8$\times 10^{-8}$ \\
		$c_0^\mathrm{L}$ & -1.4$\times 10^{-7}$ & -1.8$\times 10^{-7}$ & -1.4$\times 10^{-7}$ \\	
		\hline\hline
	\end{tabular}
	\caption{The sound velocity and quadratic coefficient, Eq. (\ref{eq:isotropic}), in the best isotropic dispersion approximation to the phonon dispersion of III-As binaries.}
	\label{tab:isodisp}
\end{table}

%


\end{document}